\documentclass[usenatbib,a4paper]{mn2e}

\usepackage{epsfig,times,amsmath,amsfonts,amssymb,multirow,array,fleqn}

\title[Improving PSF modelling for weak gravitational lensing]{Improving PSF modelling for weak gravitational lensing using new methods in model selection}

\author[Barnaby Rowe]{Barnaby~Rowe\thanks{E-mail: rowe@iap.fr} \\ 
Institut d'Astrophysique de Paris, UMR7095 CNRS, Universit\'{e}
Pierre et Marie Curie -- Paris 6, 98 bis, Boulevard Arago, 75014
Paris, France}

\begin{document}
\maketitle

\newcommand{\expect}[1]{\left\langle #1 \right\rangle} 
\newcommand{\dif}{\mbox{$\mathrm{d}$}}
\newcommand{\thetab}{\mbox{\boldmath$\theta$}}
\newcommand{\me}{\mbox{$\mathrm{e}$}}
\newcommand{\mi}{\mbox{$\mathrm{i}$}}
\newcommand{\mes}{\mbox{\scriptsize$\mathrm{e}$}}
\newcommand{\mos}{\mbox{\scriptsize$\mathrm{o}$}}
\newcommand{\mess}{\mbox{\tiny$\mathrm{e}$}}
\newcommand{\moss}{\mbox{\tiny$\mathrm{o}$}}
\newcommand{\msep}{\mbox{\scriptsize$\mathrm{sep}$}}
\newcommand{\mis}{\mbox{\scriptsize$\mathrm{i}$}}
\newcommand{\mcov}{\mbox{$\mathrm{cov}$}}
\newcommand{\mtot}{\mbox{\scriptsize$\mathrm{tot}$}}
\newcommand{\mtan}{\mbox{\scriptsize$\mathrm{tan}$}}
\newcommand{\mrot}{\mbox{\scriptsize$\mathrm{rot}$}}
\newcommand{\mprop}{\mbox{\scriptsize$\mathrm{prop}$}}
\newcommand{\mcom}{\mbox{\scriptsize$\mathrm{com}$}}
\newcommand{\mang}{\mbox{\scriptsize$\mathrm{ang}$}}
\newcommand{\mstar}{\mbox{\scriptsize$\mathrm{star}$}}
\newcommand{\mpairs}{\mbox{\scriptsize$\mathrm{pairs}$}}
\newcommand{\mMC}{\mbox{\scriptsize$\mathrm{MC}$}}
\newcommand{\mobs}{\mbox{\scriptsize$\mathrm{obs}$}}
\newcommand{\mMCs}{\mbox{\tiny$\mathrm{MC}$}}
\newcommand{\nn}{\nonumber \\}
\newcommand{\mmin}{\mbox{\scriptsize$\mathrm{min}$}}
\newcommand{\mmax}{\mbox{\scriptsize$\mathrm{max}$}}
\newcommand{\mt}{\mbox{\scriptsize$\mathrm{t}$}}
\newcommand{\mm}{\mbox{\scriptsize$\mathrm{m}$}}
\newcommand{\xb}{\mbox{\boldmath$x$}}
\newcommand{\rb}{\mbox{\boldmath$r$}}

\newcommand{\nat}{Nat}
\newcommand{\mnras}{MNRAS}
\newcommand{\apj}{ApJ}
\newcommand{\apjl}{ApJL}
\newcommand{\apjs}{ApJS}
\newcommand{\physrep}{Phys.~Rep.}
\newcommand{\aap}{A\&A}
\newcommand{\aaps}{A\&AS}
\newcommand{\aj}{AJ}

\begin{abstract}
  A simple theoretical framework for the description and interpretation of   spatially correlated modelling residuals is presented, and the resulting   tools are found to provide a useful aid to model selection in the context of   weak gravitational lensing.  The description is focused upon the specific   problem of modelling the spatial variation of a telescope point spread   function (PSF) across the instrument field of view, a crucial stage in   lensing data analysis, but the technique may be used to rank competing   models wherever data are described empirically. As such it may, with further   development, provide useful extra information when used in combination with   existing model selection techniques such as the Akaike and Bayesian   Information Criteria, or the Bayesian evidence.  Two independent diagnostic   correlation functions are described and the interpretation of these   functions demonstrated using a simulated PSF anisotropy field.  The efficacy   of these diagnostic functions as an aid to the correct choice of empirical   model is then demonstrated by analyzing results for a suite of Monte Carlo   simulations of random PSF fields with varying degrees of spatial structure,   and it is shown how the diagnostic functions can be related to requirements   for precision cosmic shear measurement.  The limitations of the technique,   and opportunities for improvements and applications to fields other than   weak gravitational lensing, are discussed.
\end{abstract}

\begin{keywords}
gravitational lensing -- methods: data analysis -- methods: statistical -- cosmology: observations -- cosmology: large-scale structure of the Universe.
\end{keywords}

\section{Introduction}\label{sect:intro}
The study of weak gravitational lensing (see, e.g.,
\citealp{schneider06} for a recent review) promises much as a means of
detecting and measuring massive structure on cosmological scales.
Through its sensitivity to all lensing mass whether baryonic or
exotic, weak lensing potentially provides direct measurement of the
cosmological matter power spectrum (e.g.\ \citealp{fuetal08}), a means
of relating this power to visible structure
(e.g.\ \citealp{hoekstraetal05,mandelbaumetal06gglens};
\citealp*{tianetal09}), and the mapping of individual clusters and
super-clusters (e.g.,
\citealp{masseyetal07nature,hoekstra07,heymansetal08}).  However,
extracting an unbiased and uncontaminated shear signal from real
telescope images in current and upcoming surveys represents an
unprecedented technical challenge.

Much work has gone into testing and refining the many competing weak
lensing data processing pipelines, using simulations of survey imaging
data with known input lensing signals
\citep{heymansetal06step,masseyetal07step,bridleetal08}.  This
simulation work has understandably concentrated upon the problem of
shear measurement from noisy galaxy images, after their convolution
with an anisotropic telescope point spread function (PSF) and
subsequent pixelization onto CCD arrays.  However, there are other
important stages in any weak lensing analysis that have been subjected
to less scrutiny, such as: stacking and dithering of exposures to
create science images, selection of stars for PSF modelling, colour
dependence of instrument PSFs when using broad band filters, and the
accurate modelling of the spatial variation of the PSF across the
telescope field of view. \citet{paulinetal08} and \citet*{paulinetal09} have
recently conducted work into the amount of information required for
characterizing typical PSFs at a given point, described in terms of
`complexity' and `sparsity', but the amount of information about the
overall spatial variation across the sky is less well-addressed by the literature.
Moreover, this spatial variation in the PSF is purposefully ignored by
current shear measurement testing simulations (e.g.,
\citealp{bridleetal08}), reserved as a separate issue. 

The work in this paper is
motivated towards finding ways to improve this aspect of PSF modelling, which
traditionally takes the form of fitting polynomial surfaces to
quantities that represent important properties of the PSF.  In the KSB
method (\citealp*{kaiseretal95}; \citealp{hoekstraetal98}) these
quantities are frequently the two components of stellar anisotropy
correction, estimated using the measured ellipticities of stars in the
field of view.  For techniques that model PSFs with shapelet basis
functions (see, e.g.,
\citealp{bernsteinjarvis02,refregier03,refregierbacon03,masseyrefregier05})
it is the spatial variation in each shapelet coefficient that is
described using parametrized surfaces.  Regardless of the quantity being modelled there is considerable freedom in the choice
of the functional form of the fitting surface (see, however,
\citealp{rhodesetal07}; \citealp*{jarvisetal08} for physically
motivated PSF models), and simple bivariate polynomials are typically
used.
Known problems with polynomial fitting surfaces, including reduced
stability at field edges and corners, have been noted but not
necessarily tackled beyond suggestions of other, perhaps better
behaved, functional schemes (e.g., \citealp*{vanwaerbekeetal05}, who
used dense stellar fields to characterize structure in the PSF).
Choices must also be made as to whether to model the PSF independently
in areas imaged by different CCD chips, and how to weight stars of
different signal to noise in the same fit.  Unfortunately there is
often little guidance from the data itself, as accurately estimating
the uncertainty on a stellar ellipticity measurement or shapelet
coefficient is difficult.

One recent development has been the suggestion of Principal Component
Analysis (PCA) as a means of building up knowledge of the PSF
\citep{jarvisjain04,schrabbacketal09}, allowing the use of more complex polynomial
surfaces coupled with a more rigourous quantification of the degree of
information redundancy. The technique uses a large number of images to
explore the principle vectors in the space of observed PSF models, and
will clearly form a crucial part of future PSF modelling for large
surveys.  Using PCA, overfitting can be controlled whilst ensuring that
all the observable features in the PSF are properly modelled. However,
this approach has one important caveat: it assumes that there is no
independent random or complex quasi-random component to the PSF
anisotropy in any given field. This assumption will conceivably be
broken for ground-based data (possibly even from space), specifically
if the anisotropy is a combination of predictable \emph{and} complex
or chaotic effects. In this paper the investigation will instead focus
upon a complementary question: whether there are further tests of the
modelling quality of a single PSF anisotropy map, including those created using PCA, without requiring it be drawn from a physically predictable underlying distribution.

Tests for overall control of systematics, and indirectly therefore the
quality of the PSF model, do exist in weak lensing once the shear
signal can be decomposed into E-mode/B-mode components (e.g.,
\citealp{crittendenetal02,schneideretal02bmode}).  However, these tests are
only possible \emph{after} the lensing shear is measured, at the end
of the analysis, and B-modes may also be generated by intrinsic
alignments and source clustering.  An important investigation into the
effects of imperfect PSF modelling was performed by
\citet{hoekstra04}, who analyzed residual correlations in PSF models
using the aperture-mass statistic and studied the impact of these
correlations upon cosmic shear measurements.  This paper naturally
follows on from that work by providing a formal discussion of the
reasons for residual correlations in poorly modelled data.  In
addition, it presents a first investigation into whether such
correlations may be used as an aid to the systematic selection of PSF
modelling schemes, and as an aid to modelling in general.

The assessment of goodness of fit (see, e.g., \citealp{lupton93}) of a
given model to the physical data is a vital stage in any scientific
analysis, and the related field of model selection is attracting
increasing interest within astronomy as a means of evaluating evidence
for competing cosmological models (for recent reviews in astrophysical
contexts see \citealp*{liddleetal06} and \citealp{trotta08}; for
recent discussions and applications see, e.g.,
\citealp{liddle07,efstathiou08,kurekszydlowski08}).  Estimates and
uncertainties upon model parameters derived from any fit are
meaningless if the model itself is an unlikely match to the data, and
if further analyses depend upon the accuracy or stability of this
model then later conclusions may be biased or subject to unnecessary
additional variance \citep{paulinetal09}. The problem can become acute
in applications where empirical or unverified physical models must be
used, or where errors upon measured data points are difficult to
estimate accurately.  These problems are precisely those
encountered when attempting to model the spatial variation of the PSF
in weak lensing applications.

The most famous and often-used diagnostic of goodness of fit is the
chi-squared statistic (see \citealp{lupton93}), and simple chi-squared
per degree of freedom arguments are often used as a means of model
selection (see, e.g., \citealp{spergeletal07}).  A related measure
that generalizes to non-Gaussian distributions is the Akaike
Information Criterion (AIC), derived from an approximate minimization
of the Kullback-Leibler information entropy (see, e.g.,
\citealp{liddleetal06,trotta08}).  Bayesian probability theory (see
\citealp{gelmanetal03} for a comprehensive general reference) also
provides two further guides to model selection: the full calculation
of the Bayesian evidence, and its related approximation the Bayesian
Information Criterion (BIC: again see \citealp{trotta08}).  These
criteria all use calculations of the statistical likelihood
$\mathcal{L}$ of a dataset given the model in question, either the via
integration of $\mathcal{L}$ over the full possible parameter space in
the case of the Bayesian evidence, or by comparison of the
best-fitting model maximum likelihood $\mathcal{L}_{\mmax}$ to the
number of parameters in the model.  In order to reliably calculate
these `data-given-model' likelihoods, the probability distributions
$p(y_i)$ of individual measured data points $y_i$ must be known or
well-approximated; as discussed above, this is seldom the case in PSF
modelling contexts.  

One important topic of this paper is to discuss
other properties of the relationship between model and data that can
be usefully explored without good prior knowledge of the uncertainties
upon individual data points.
A related investigation, merely initiated by this work,
will be to begin to understand what information is being lost when
employing model selection arguments based entirely upon
data-given-model likelihoods or related criteria. 
As will be shown, such
information may be of use when diagnosing goodness of fit. The spatial
correlation of residuals for both underfitting and overfitting models
is partly predictable, and this insight can also be used to guide
modelling improvements in a way that conforms to the principle of
Occam's Razor.

The structure of this paper is as follows: in Section \ref{sec:theory}
a basic theory of correlations in model residuals is described,
specifically aimed at models of stellar ellipticity. This leads to the
construction of two independent diagnostic functions, and makes
predictions for the behaviour of these functions for under- and
overfitting models. Section \ref{sect:testcase} then applies these
diagnostics to a test-case scenario of a simulated starfield with a
known underlying PSF anisotropy model.  In Sections \ref{sect:monte} \& \ref{sect:arbitrary}
these tests are repeated on a suite of simulated
starfields with varying degrees of spatial structure in the PSF
map. Relating the diagnostics to requirements for cosmic shear surveys is discussed in Section \ref{sect:relation}.  In Section \ref{sect:extension} the possibility of a
generalized extension beyond weak lensing is discussed, followed in Section \ref{sect:conc} by
a general summary and conclusions.

\section{Spatial correlations in model residuals}\label{sec:theory}
\subsection{Basic theory and assumptions}\label{sec:basic}
The starting point in this analysis is to construct a simplified
description of the process of modelling PSF anisotropy across the
$x$-$y$ plane, such as across a telescope field of view.  In what follows the
discussion is limited to models of complex, spin-2 pseudo-vector 
fields (i.e.\
$f(x,y) = |f|(x,y) \me^{\mis 2 \theta(x,y)}$) on a 2D plane, 
but such arguments may be generalized to scalar or spin-$n$ fields,
and to other spatial dimensionalities (see Section \ref{sect:conc}).

An observed complex ellipticity field is described as the sum of two
contributions, the unknown `true'
ellipticity field $e_{\mt}$ and $N$, a stochastic complex variable
describing noise upon ellipticity measurements. This field is labelled
$e(x,y) = e_1 + \mi e_2$, where:
\begin{equation}\label{eq:e}
e = e_{\mt} + N.
\end{equation}
In almost all that follows the explicit $x$-$y$ dependence of fields such as
$e_{\mt}$ will be dropped for brevity, and should be implied.
The description of discrete observations $e_i$ by a 
continuous `quasi-field' $e$ is also a notational convenience: 
measured quantities will always be an $e(x_i,y_i)$ sample of the 
proposed field $e$.  We assume that the stochastic
noise term satisfies $\expect{N} = 0$.

If a best fit parametrized model is then applied to describe $e$, 
one can choose to write the modelled ellipticity $e_{\mm}$ as
\begin{equation}\label{eq:em}
e_{\mm} = e_{\mt} + m(e_{\mt}, N ; f_{\mm}),
\end{equation}
where $m = e - e_{\mt}$ will hereafter be referred to as the 
\emph{inaccuracy} in the model, also unknown,
and $f_{\mm}$ is simply a label denoting the functional scheme used
to fit the spatial variation of $e$ (e.g.\ a second order bivariate
polynomial).  This expression makes explicit the dependence of the
inaccuracy $m$ upon $e_{\mt}$, the ensemble of 
discrete realizations of $N$, and $f_{\mm}$. In general this 
dependence will be non-trivial to describe, but the aim of this paper is to look for diagnostic tests by which the functional properties of $m$ can be constrained.

The principal tool in this work is the two point ellipticity
correlation function, the observable quantity used to extract signal
in cosmic shear studies. In this the investigation follows
\citet{hoekstra04}, as well as the later work of
\citet{vanwaerbekeetal05} and \citet{hoekstraetal06}, in analyzing
correlations in corrected PSF patterns as a means of testing
anisotropy models.  Specifically the $\xi_{\pm}(r)$ correlation
functions will be used, defined as
\begin{equation}\label{eq:xipm}
\xi_{\pm} (r) = \expect{e_{\mtan}(\xb + \rb) e_{\mtan}(\xb)}
 \pm \expect{e_{\times}(\xb + \rb) e_{\times}(\xb)},
\end{equation}
where $e_{\mtan}$ and $e_{\times}$ are the known as the
\emph{tangential} and \emph{rotated} components of the ellipticity,
and the angle brackets denote an averaging over all \emph{pairs} of
points separated by a distance $r$. It should be noted therefore that
this definition implicitly averages over all angles, which is not a
problem in the cosmic shear case where correlations are assumed
isotropic so that $\xi(\rb) = \xi(r)$;
if the field is not strictly isotropic then it must be borne in mind
that $\xi(r)$ is the angular average correlation, and thus that some
information may have been lost. 

The tangential and rotated components
are defined for the complex ellipticities $e$ of each pair as
\begin{equation}\label{eq:tanrot}
e_{\mtan} + \mi e_{\times} = - \me^{-2\mis \phi}(e_1 + \mi e_2)
\end{equation}
(see, e.g., \citealp{schneider06}), where $\phi$ is the angle between the abscissa and the line joining the location of each member of the pair of points.  With a discrete number of ellipticity measurements the quantities in equation \eqref{eq:xipm} can then be estimated by
\begin{equation}\label{eq:xipprac}
\xi_{\pm}(r) = \frac{1}{N_{\mpairs}} \sum_{\mpairs} e_{\mtan}(\xb +
\rb) e_{\mtan}(\xb) \pm e_{\times}(\xb + \rb) e_{\times}(\xb)
\end{equation}
in finite bins of $r$. Such correlation function estimates, when made
using PSF-corrected galaxy ellipticities, are the primary observable
quantity in modern studies of cosmological weak lensing (see, e.g.,
\citealp{fuetal08,schrabbacketal07}).

In the following discussion, which focuses upon $\xi_+$ as a diagnostic of PSF modelling, frequent use will be made of a useful shorthand notation:
\begin{eqnarray}
\expect{e^* e} \!\! & \equiv & \expect{(e_{\mtan} - \mi e_{\times})(e_{\mtan} + \mi e_{\times})} \label{eq:auto} \\
& = & \xi_+(r)  \nn 
 & + & \mi \expect{ e_{\mtan}(\xb + \rb) e_{\times}(\xb) - e_{\times}(\xb + \rb) e_{\mtan}(\xb)}.  \label{eq:xpart}
\end{eqnarray}
Due to parity symmetry (see \citealp{schneider06}) the imaginary
second term in equation \eqref{eq:xpart} tends to zero, and so
$\expect{e^*e}(r) = \xi_+(r)$. Similarly, $\expect{e^*_{\mm}
  e_{\mm}}(r)$ will be used to denote the $\xi_+(r)$ autocorrelation
in the model $e_{\mm}$.  This notation is convenient not only as a
labelling convention but also as a computational tool, and so will be
used exclusively hereafter.

In order to proceed, two simplifying assumptions are made about the noise. Firstly, it is assumed that $N$ is spatially \emph{un}correlated so that 
\begin{equation}\label{eq:noncor}
\expect{N^*N}(r) =  0
\end{equation}
for all $r > 0$.  Secondly, it is assumed
that the cross-correlation between $N$ and the unknown
$e_{\mt}$ is also zero, so that
\begin{equation} \label{eq:nonecor}
\expect{ e^*_{\mt} N + N^* e_{\mt} }(r) = 0.
\end{equation}
Situations in which equations \eqref{eq:noncor} and \eqref{eq:nonecor} no
longer hold can be envisaged, such as in the presence of problems with
reduced Charge Transfer Efficiency (CTE) in telescope CCDs (see,
e.g.\, \citealp{rhodesetal07}) or preferences for certain PSF directions due to large pixel sizes or under sampled stellar images. In this theoretical analysis these factors are assumed to be small, but if necessary the assumption may be easily tested and corrected for using simulated data.

\subsection{Fit diagnostics}\label{sec:diagnostics}
For an ideal fit $m=0$ everywhere, but this is unrealistic in the case
of noisy data and finite numbers of observations. For the case of an
imperfect but well-constrained, stable and accurately predictive model
the following three conditions should be simultaneously fulfilled:
\begin{eqnarray}
  \expect{m^* e_{\mt} + e^*_{\mt} m }(r) & \label{eq:c1} \simeq & 0 \\
  \expect{m^*N + N^* m}(r) & \simeq & 0 \label{eq:c2} \\
  \expect{m^*m}(r) & \simeq & 0 \label{eq:c3}
\end{eqnarray}
for all $r > 0$.  These conditions will be met if the
modelling inaccuracies $m$ can, like the noise, 
be approximately described as an independent, stochastic variable $m = M$ with $\expect{M} = 0$.

The first of these conditions \eqref{eq:c1} requires that the
inaccuracies $m$ should be distributed at random with respect to the
true ellipticity distribution, and thus that the cross-correlation
between these quantities be zero; this condition will be broken if the
model is systematically \emph{underfitting} the true ellipticities
$e_{\mt}$, as will be discussed in Section \ref{sec:under} below.  The
second condition \eqref{eq:c2} explicitly states that $m$ is required
to be uncorrelated with the noise $N$, expected if the scheme
$f_{\mm}$ allows significant \emph{overfitting} of the data.  The
third and final condition \eqref{eq:c3} requires that $m$ show no
spatial autocorrelation, which will be fulfilled if the inaccuracies
at all points are mutually randomly distributed. This third condition
may be broken for both overfitting, where neighbouring $m$ values may
be correlated to the extent allowed by instabilities in the chosen
$f_{\mm}$, and underfitting, where $e_{\mm}$ systematically fails to
reproduce observable features in $e_{\mt}$.

Unfortunately, the functions in equations
\eqref{eq:c1}-\eqref{eq:c3} cannot be measured
directly, as $m$, $N$ and $e_{\mt}$ are unknown. 
However, it is possible to construct two independent,
observable correlation functions 
with which to attempt to constrain these quantities. 
There is freedom in how these are
defined (as will be discussed below), 
but the following useful forms are suggested: 
\begin{eqnarray}
D_1(r) & \equiv & \expect{(e- e_{\mm})^*(e- e_{\mm})}(r) \label{eq:d1} \\
D_2(r) & \equiv & \expect{e^* (e - e_{\mm}) + (e - e_{\mm})^* e }(r) \label{eq:d2}.
\end{eqnarray}
These two diagnostic 
quantities $D_1$ and $D_2$ 
may be easily estimated from modelled stellar fields
using routines that are standard in statistical lensing.  Using the
definitions and assumptions described by equations
\eqref{eq:e}-\eqref{eq:nonecor}, they can be expressed in terms
of the unknown quantities as follows:
\begin{eqnarray}
D_1(r) & = & - \expect{m^* N + N^*m}(r) + \expect{m^*m}(r) \label{eq:dd1}\\
D_2(r) & = & - \expect{m^*N + N^*m}(r) - \expect{m^* e_{\mt} + e_{\mt}^*m}(r)
\label{eq:dd2} .
\end{eqnarray}
It is noted immediately that there are only two quantities with which to
constrain the three unknowns of conditions \eqref{eq:c1}-\eqref{eq:c3}, and
the system is therefore underdetermined.  This is a natural
consequence of there being only two directly 
observable quantities ($e$ and $e_{\mm}$) with which to place
constraints upon correlations in $m$, $N$ and $e_{\mt}$.
The system of equations described by \eqref{eq:dd1} and \eqref{eq:dd2} cannot therefore be solved to demonstrate any of
\eqref{eq:c1}-\eqref{eq:c3} uniquely. Instead, one 
may determine only a general family of solutions. In vector notation this
solution is simply
\begin{equation}\label{eq:gensol}
\left( \begin{array}{c} \expect{m^*m}(r) \\
\expect{m^* N + N^*m}(r) \\
\expect{m^* e_{\mt} + e^*_{\mt}m}(r) \end{array} \right)
= \left( \begin{array}{c} 1 \\ 1 \\ -1 \end{array} \right) t
+ \left( \begin{array}{c} D_1(r) \\ 0 \\ -D_2(r) \end{array} \right)
\end{equation}
where $t$ is any real number.  Attempts to construct a third independent
observable with which to break this degeneracy do not succeed: for
example, the observable function $\expect{e^*e} - \expect{e^*_{\mm}e_{\mm}}$
may be written simply as $D_2 - D_1$.


The fact that this system of equations is underdetermined
means the $D_1(r)$ or $D_2(r)$ diagnostics can never be used to
\emph{prove} that $\expect{m^*m} = \expect{m^*N + N^*m} = \expect{m^*
  e_{\mt} + e^*_{\mt}m} = 0$ at all scales.  However, they allow the positive diagnosis of poor modelling if either are 
nonzero at a significant level. 
Moreover, in order to be erroneously led into the belief that a given model fit is accurate and
stable would require that
\begin{equation}
\expect{m^* m} \simeq \expect{m^*N + N^* m} \simeq - \expect{m^*
  e_{\mt} + e^*_{\mt}m}
\end{equation}
for all scales $r$, constituting a significant coincidence and very poor
luck.  
Finally, although the general expression in
\eqref{eq:gensol} allows an infinite family of solutions to our three
quantities (\ref{eq:c1}-\ref{eq:c3}), this solution is not the only information we have.   Further assumptions and simple
reasoning may be employed so as to predict what might be expected 
for these diagnostic measures in the cases of over- or underfitting; 
this in turn may allow the tuning of modelling schemes so as 
to better represent the data without fitting noise.  These reasoned
expectations for the general form of $D_1(r)$ and $D_2(r)$ will now be outlined.

\subsection{The case of underfitting}\label{sec:under}
In order to describe what might be expected in the case of
underfitting, we consider a hypothetical scheme $f_{\mm}$ for which the model
ellipticities $e_{\mm}$ do not adequately represent the underlying
field $e_{\mt}$.  This can be expressed in terms of the simple
model constructed in Section \ref{sec:basic}, and the further assumption that in the limit of \emph{severe} underfitting it may be approximated that
\begin{equation}\label{eq:munder}
m = m(N, e_{\mt}; f_{\mm} ) \simeq m(e_{\mt}; f_{\mm} ).
\end{equation}
The validity of this approximation will depend strongly upon the
degree to which a given fit fails to reproduce observable features in
$e_{\mt}$, but it can be motivated by an
illustrative example: a flat model
scheme $f_{\mm}$ for which $e_{\mm} = \expect{e} =$ constant being used to describe an $e_{\mt}$ which varies significantly with $x$ and $y$.  
In this case $m = \expect{e} - e_{\mt}$, which is a clear function of
$e_{\mt}$ and only very weakly dependent upon the noise via $\expect{e}$.

Given the assumption of equation \eqref{eq:munder}, it should be
expected that $m$ is approximately uncorrelated with the noise $N$ for
underfitting models (carrying only a weak dependence upon it), and thus that
\begin{eqnarray}\label{eq:nmzero}
\expect{m^* N + N^* m}(r) & \simeq & \expect{m^*(e_{\mt}; f_{\mm}) N +
  N^* m(e_{\mt}; f_{\mm})} \\
& \simeq & 0.
\end{eqnarray}
The condition expressed in equation \eqref{eq:c2} is therefore approximately fulfilled for underfitting models, meaning that observations of $D_1(r)$ and $D_2(r)$ will provide
constraints upon the quantities in equations \eqref{eq:c1} and \eqref{eq:c3}.
Using equations \eqref{eq:d1} and \eqref{eq:d2}, this then gives
\begin{eqnarray}
D_1(r) & \simeq & \expect{m^*m}(r) \label{eq:d1under} \\
D_2(r) & \simeq &  - \expect{m^* e_{\mt} + e_{\mt}^* m}(r) \label{eq:d2under} .
\end{eqnarray}
These predictions can now be used to make further conclusions as to the
expected form of these functions, allowing the positive diagnosis of
underfitting models if these expected forms are observed.

\begin{figure}
\begin{center}
\psfig{figure=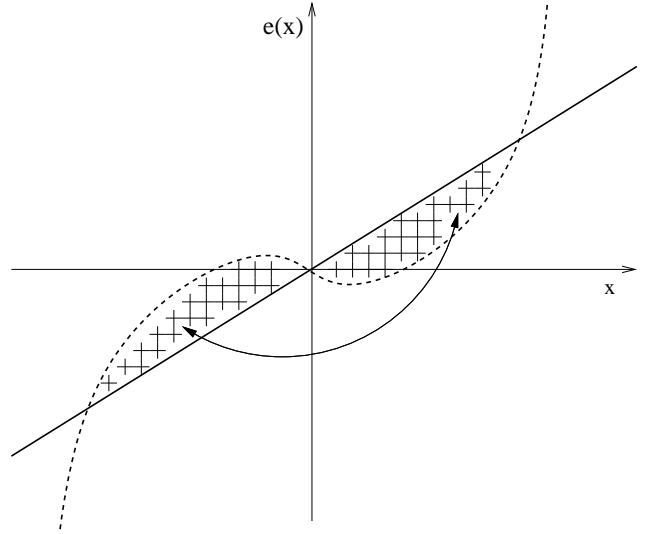,width=8.4cm,angle=0}
\caption{\label{fig:manti} Schematic showing
    how $\expect{m^*m}(r)$ may be expected to be both negative and
    positive, depending upon scale $r$, 
 for underfitting models.  The dotted line is a given `true'
 model, and the thick solid line a best-fit straight line.  The
 model inaccuracy $m$ will be correlated within the two cross-containing
 regions, but \emph{anticorrelated} on the larger scales between the
 two regions, indicated by the closed arrows.}
\end{center}
\end{figure}
From equation \eqref{eq:munder} it is clear that the form taken by
$\expect{m^*m}$ will be related to that of 
$\expect{e^*_{\mt} e_{\mt}}$, and if the underfitting is
significant then $\expect{m^*m}$ is likely to be nonzero on or
near the scales for which $\expect{e^*_{\mt} e_{\mt}}$ is nonzero.
Figure \ref{fig:manti} shows a simple example of how both
correlation and anticorrelation might be expected in underfitting
model residuals; in this simple example of a one-dimensional
underfitting model the residuals within the cross-containing regions
will be positively correlated, but \emph{between} these regions the
residuals will be negatively correlated.
Turning to the simple $e_{\mm} =
\expect{e}$ model for another illustration, using equations 
\eqref{eq:d1under} and $\expect{N} \simeq 0$ it may be written that
$D_1(r) \simeq  \expect{e_{\mt}^* e_{\mt}} -
\left|\expect{e_{\mt}} \right|^2$.  Therefore, this $D_1(r)$ may potentially be both positive or negative for differing 
ranges $r$, depending upon the functional form of $e_{\mt}$.

This behaviour will also be expected from the 
second measurable quantity $D_2(r)$, which from equation
\eqref{eq:d2under} will also be
activated by underfitting: $\expect{m^*e_{\mt} + e^*_{\mt}m}$ 
will often be nonzero for $m \simeq m(e_{\mt};f_{\mm})$.
The function $D_2(r)$ may be rewritten by 
considering that $m = e_{\mm}- e_{\mt}$, giving
\begin{eqnarray}
D_2(r) & \simeq & -\expect{m^* e_{\mt} + e^*_{\mt}m} \nonumber \\
   &=  &2\expect{e^*_{\mt}e_{\mt}} - \expect{e_{\mm}^*e_{\mt} + e_{\mt}^*e_{\mm}}.
\end{eqnarray}
Once again, $D_2(r)$ may be expected to be either
positive or negative depending upon scale $r$ and the sign of $\expect{e^*_{\mt}e_{\mt}}$.
However, the precise dependence of $D_2(r)$ may differ from that of $D_1(r)$ in a way that is difficult to predict without detailed knowledge of the $e_{\mt}$ field.

\subsection{The case of overfitting}\label{sec:over}
The case of overfitting is now considered so as to make predictions
for the behaviour of $D_1(r)$ and $D_2(r)$: if the behaviour differs
from the underfitting case this will allow the two cases to be
distinguished and positively identified, and allow subsequent
modelling improvements to be made in a guided fashion.  As mentioned
in Section \ref{sect:intro}, the application of PCA to PSF modelling
\citep{jarvisjain04} also brings control over overfitting
via the removal of identified low-importance principal components.
However, it may be that a hybrid of PCA and the simultaneous fitting
of an independent surface is necessary to account for random changes
in the PSF pattern between exposures, and so a diagnosis of possible
overfitting in this extra surface will still be desirable.  Moreover,
the behaviour of $D_1(r)$ and $D_2(r)$ in overfitting models is an
interesting investigation in itself, as the technique may be useful in
other situations in which PCA is not directly applicable (see Section
\ref{sect:extension}).

In Section \ref{sec:under} the inaccuracy $m$ was approximated as
being a function of $e_{\mt}$ and $f_{\mm}$ only.  A similar approximation 
can be argued for the opposing case of severe overfitting:
\begin{equation}\label{eq:mover}
m = m(N, e_{\mt}; f_{\mm}) \simeq m(N; f_{\mm}).
\end{equation}
This statement can be justified by considering what is meant by an
overfitting model: one which captures not only an observable physical
trend, but is also unjustifiably sensitive to random noise upon
measurements.  Such models will not in general be 
biased in a way that can be related to $e_{\mt}$, but will prove to be
unstable with respect to changes in $N$ (this concept is also
discussed in \citealp{paulinetal09}).  The
correlation properties of $m$ are now largely decided by this random
quantity and inherent correlations caused by the best-fitting 
chosen model $f_{\mm}$.
Taking this assumption, and using equations \eqref{eq:mover} and 
\eqref{eq:nonecor}, leads to
\begin{eqnarray}
\expect{m^* e_{\mt} + e_{\mt}^* m }(r) & \simeq &  
\expect{m^*(N; f_{\mm}) e_{\mt} + e^*_{\mt} m(N; f_{\mm}) }  \\
 & \simeq & 0,\label{eq:nomecor}
\end{eqnarray}
via the same reasoning as equation \eqref{eq:nmzero}.  This corresponds to
saying that condition \eqref{eq:c1} will be fulfilled
when severely overfitting data, just as in Section \ref{sec:under} it was
argued that condition \eqref{eq:c2} is automatically fulfilled if data
is being underfit.

Using this result, the effects of overfitting upon the forms of
$D_1(r)$ and $D_2(r)$ can be predicted.  From equations \eqref{eq:d1},
\eqref{eq:d2}, and \eqref{eq:nomecor} it may be
written that
\begin{eqnarray}
D_1(r) & = & \expect{m^*m}(r) - \expect{m^*N + N^*m}(r) \label{eq:d1over} \\
D_2(r) & \simeq & -\expect{m^* N + N^* m}(r) \label{eq:d2over} .
\end{eqnarray}
It is noted immediately that the expression for $D_1(r)$
remains unsimplified: unlike $\expect{m^* e_{\mt} + e^*_{\mt} m}$, the
$\expect{m^*m}$ term will not necessarily vanish for overfitting models,
despite the assumption of uncorrelated noise in equation \eqref{eq:noncor}.
As can be seen from equation \eqref{eq:mover}, there remains the
possibility of correlation due to the inherent properties of the functional form of the model $f_{\mm}$ used to unstably fit the data.

It is expected that the condition 
\eqref{eq:c3} is \emph{not} fulfilled when we are
overfitting the data, and thus that $D_2(r)$ will be nonzero.  It can furthermore be said that
a \emph{positive} cross-correlation will be expected between $N$ and
$m$, so that
\begin{equation}\label{eq:mncor}
\expect{m^*N + N^* m} > 0.
\end{equation}
This can be justified by considering once again $m = e_{\mm} -
e_{\mt}$, which combined with equation \eqref{eq:nonecor} then gives 
\begin{equation}\label{eq:mnneg}
\expect{m^*N + N^* m}(r) = \expect{e_{\mm}^*N + N^* e_{\mm}}(r) > 0.
\end{equation}
This last inequality indeed expresses a definition of
overfitting, being that the best-fitting model $e_{\mm}$ shows some
significant average correlation with the particular realization
of the noise $N$. It is then clear from
equations \eqref{eq:d2over} and \eqref{eq:mnneg} that 
\begin{equation}
D_2(r) = -\expect{m^* N + N^* m}(r) < 0,
\end{equation}
i.e.\ nowhere positive and potentially significantly negative, for overfitting models.
In Section \ref{sec:under} it was seen that
$D_2(r)$ is expected to be positive over some range of $r$ for
severely underfitting models.
Whilst the minimization of $D_2(r)$ will lead to an
optimal fit in any case, this difference in behaviour between under-
and overfitting offers hope of diagnosing successfully between the
two cases, allowing informed improvements in modelling to be made at
each stage.

Furthermore, it can be argued that in most cases the
$\expect{m^*m}$ due to undue freedom in the model $f_{\mm}$ 
will be small compared to $\expect{m^* N + N^*m}$; if this is so
the $D_1(r)$ function may also be used to distinguish under- and overfitting. The argument relies on the following insight: an overfitting model, but one
that has nevertheless been constructed by minimizing deviations from
the observed data, will in all but the most pathological 
cases be expected to have inaccuracies $|m(N; f_{\mm})|
\le |N_i|$ at each point $(x_i, y_i)$.  The limiting example of this
behaviour is an `ultimate overfit', 
for which the model is simply $e_{\mm} = e$ (and
thus $m = N$), with an interpolation or spline between stellar data points. Such a model takes no account of the fact that there may be noisy or imperfect measurements among the input data. 
Using $|m(N; f_{\mm})| \le |N|$ then gives
\begin{equation}\label{eq:d1dodgy}
\expect{m^* m} \le \frac{1}{2} \expect{m^* N + N^*m} < 
\expect{m^* N + N^*m}
\end{equation}
which in turn implies that
\begin{equation}
D_1(r) = \expect{m^*m}(r) - \expect{m^*N + N^*m}(r) < 0
\end{equation}
in the same way as $D_2(r)$. This offers further hope of positive, distinguishable diagnoses of both under- and overfitting, based on whether $D_1(r)$ and $D_2(r)$ are observed to be both positive and negative, or negative only, respectively.

To summarize, the functional behaviour of the $D_1(r)$ and
$D_2(r)$ diagnostic functions for both over- and underfitting
modelling schemes $f_{\mm}$ has been predicted using simple 
assumptions about the nature of the modelling, the noise, and the 
data themselves.  The most important, and
potentially least secure, of these assumptions are
the validity of the approximate functional dependencies of $m$,
given in equations \eqref{eq:munder} and \eqref{eq:mover}.  In order
to test these points of reasoning, and to explore
the strength of each diagnostic function in a realistic modelling
situation, the following Section will test the efficacy of $D_1(r)$
and $D_2(r)$ using a simulated PSF anisotropy map with a known, underlying
ellipticity model.

\section{Tests on a simulated anisotropy map}\label{sect:testcase}

The diagnostic functions $D_1(r)$ and $D_2(r)$ have been shown to offer some promise as diagnostics of both over- and underfitting models.  The discussion has, from the start, focused upon the modelling of ellipticity fields across a plane, a process very commonly undertaken in the modelling of anisotropic PSFs for weak lensing (e.g., \citealp{kaiseretal95,hoekstraetal98,leauthaudetal07,fuetal08}).  Improving this modelling is the primary motivation of this work, and so in order to test $D_1(r)$ and $D_2(r)$ it is appropriate to simulate modelling conditions similar to those encountered in such weak lensing analyses, exploring whether some of the simple insights of \ref{sec:theory} hold validity in such a regime.  In this Section the discussion will concentrate upon a single simulated PSF anisotropy field (hereafter referred to as simply the starfield) in order to illustrate the forms of $D_1(r)$ and $D_2(r)$ in some detail, and give visual examples.

\subsection{Constructing the starfield}\label{sect:starf}

The simulated starfield is designed to mimic measurements of
ellipticity from 2500 stars across a square telescope field of view of
area 1 deg$^2$.  The
number of stars in the field, the field shape, typical PSF anisotropy
and measurement noise are loosely based upon lensing observations from
the Canada-France-Hawaii Telescope Legacy Survey-Wide (CFHTLS-W, e.g.\ \citealp{fuetal08,hoekstraetal06}).  An ellipticity
measurement $e$ was assigned to each star as modelled in equation
\eqref{eq:e}, consisting of a true underlying model $e_{\mt}$ and
additive noise term $N$, each scaled so as to resemble the CFHTLS-W.

\begin{figure}
\begin{center}
\psfig{figure=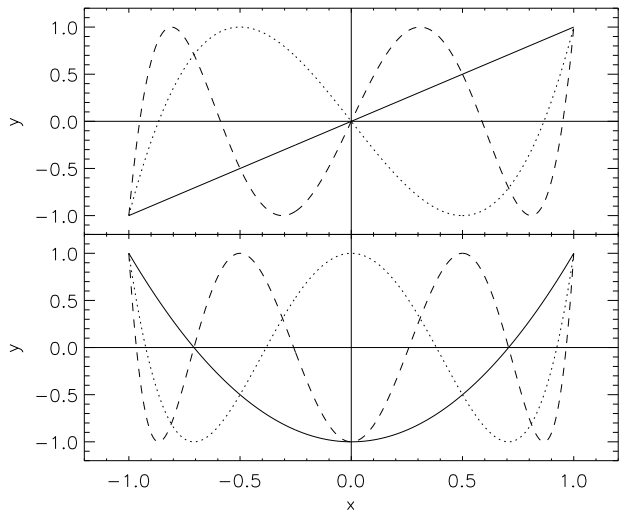,width=8.4cm,angle=0}
\caption{\label{fig:cheby} Lowest order Chebyshev polynomials of the first kind,
  plotted in the interval $x \in$ [-1,1]. Upper panel: $T_1(x)$ (solid
  line), $T_3(x)$ (dotted line), $T_5(x)$ (dashed line). Lower panel:
  $T_2(x)$ (solid line), $T_4(x)$ (dotted line), $T_6(x)$ (dashed
  line). The normalization properties of these functions on the 
chosen interval make them well-suited for simulating $e_{\mt}$ fields with
approximately equal degrees of spatial structure on varying scales.}
\end{center}
\vspace*{-1mm}
\end{figure}
In order to construct $e_{\mt}(x,y)$ the telescope field of view is
first defined upon a set of coordinates $x', y'$, with both $x'$ and
$y'$ varying in the interval [-1,1]. Each component of the true field
$e_{\mt}$ was then modelled by a bivariate, fifth order Chebyshev polynomial
defined as
\begin{equation}\label{eq:ai}
(e_{\mt})_i = \sum_{j,k=0}^{j+k \le 5} a_{ijk} T_j(x') T_k(y')
\end{equation}
with $i=1,2$ denoting the real and imaginary parts of $e_{\mt}$
respectively, and where $T_j(x)$ is the $j$th order Chebyshev
polynomial of the first kind (see
Figure \ref{fig:cheby}; also \citealp{arfkenweber05}). From this Figure the
reason for choosing these Chebyshev polynomials becomes
clear: each term in equation \eqref{eq:ai}
will add contributions of similar magnitude across the interval [-1,1].
The value of each $a_{ijk}$ was then assigned by random sampling from
a Gaussian distribution of mean $\mu_a =0$ and standard deviation $\sigma_a = 0.01$. The
coordinates are then transformed from $x',y'$ to an `observed' 
$x,y$ (in arcminutes)
via
\begin{equation}
x' = \frac{2x}{60} - 1, \:\:\:\:\:\:\: y' = \frac{2y}{60} - 1,
\end{equation}
leading to a functional expression $e_{\mt}(x,y)$ across the $60 \times 60$ arcmin$^2$ simulated field of view.

\begin{figure}
\begin{center}
\vspace*{-2.5mm}
\psfig{figure=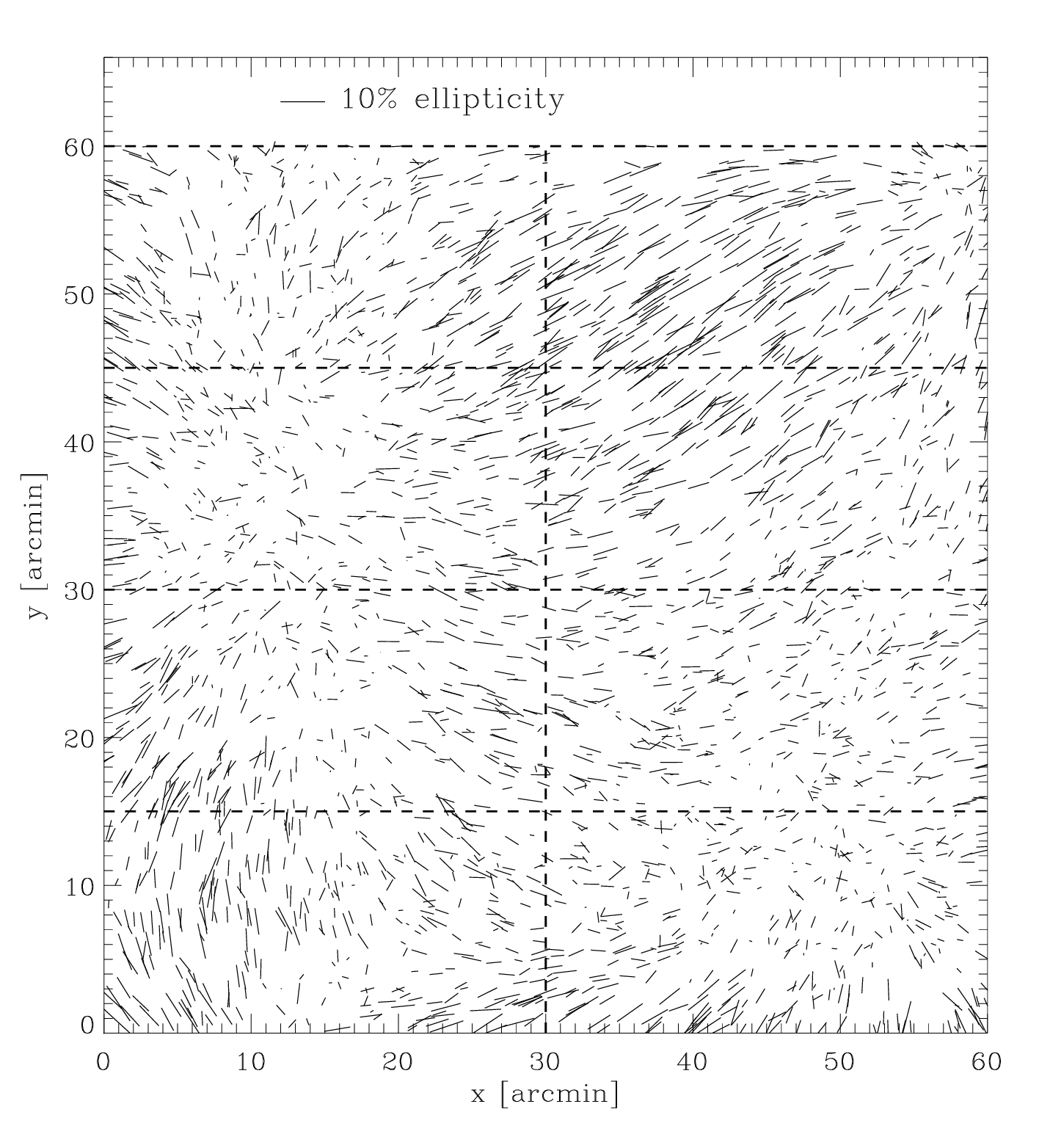,width=8.4cm,angle=0}
\psfig{figure=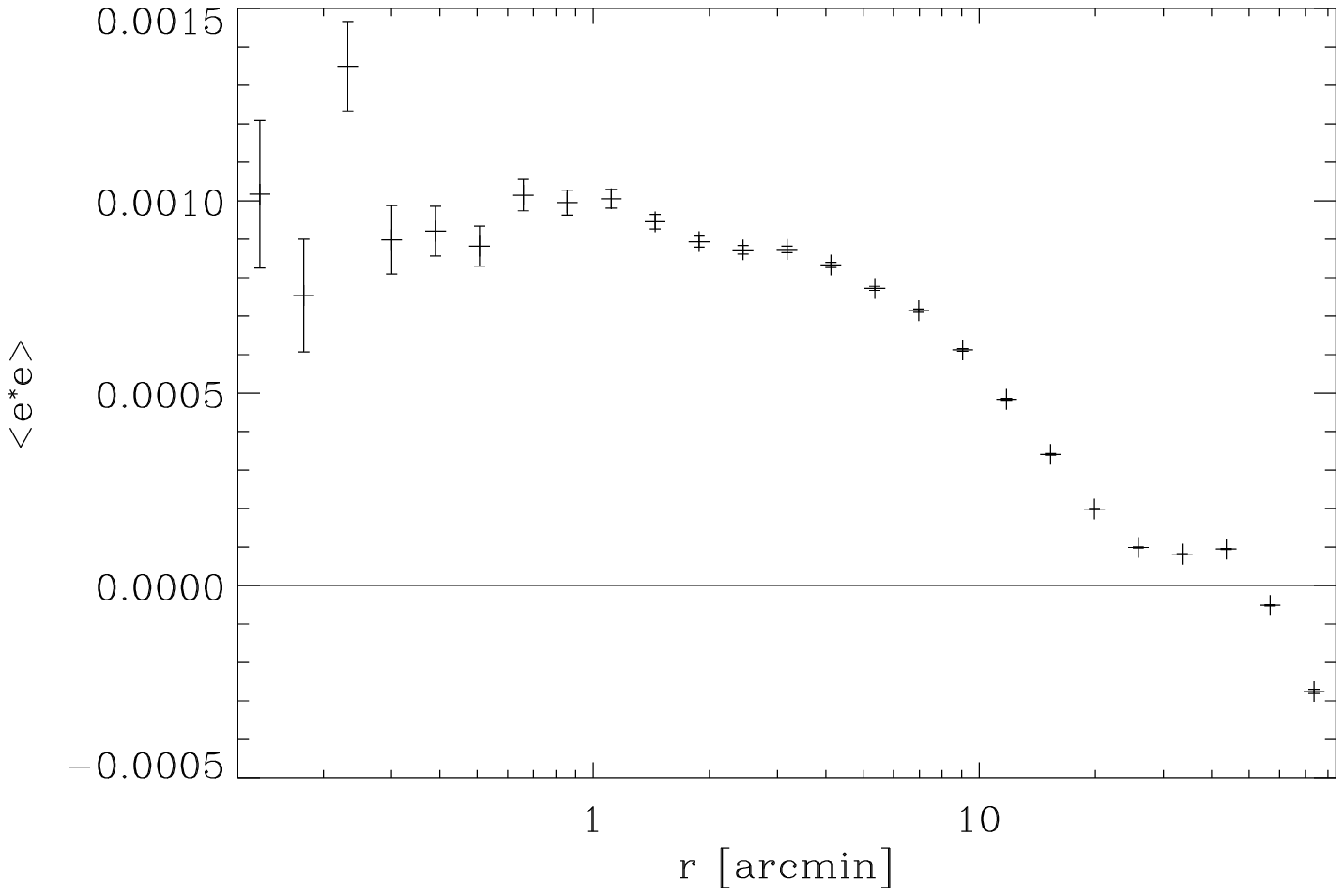,width=8.4cm,angle=0}
\caption{\label{fig:whisker} Upper panel (a): 
Whisker plot showing the positions and
  ellipticities of the 2500 stars in the simulated 1 deg$^2$ starfield, as described in Section \ref{sect:starf}. The simulated noise on each component of `measured' ellipticity $e$ 
is $\sigma_N = 0.015$. The dashed lines show the boundaries of the 8 chips
defined when performing an artificial `chipwise' fit to the simulated data.
Lower panel (b): Autocorrelation $\expect{e^*e}$ in the same simulated starfield.}
\end{center}
\end{figure}
The random noise $N$ added to each component of $e_{\mt}$ was then
modelled as a Gaussian random variable with mean $\mu_N =0$ and
standard deviation $\sigma_N = 0.015$, matching scatter in ellipticity
measurements for the bright stars (typically with $i$-band magnitude
$i < 22$) used in PSF modelling in the CFHTLS-W.  The resulting simulated
ellipticities for this test field can be seen in Figure
\ref{fig:whisker}, along with the $\expect{e^*e}$ correlation function
for the starfield.

It should be noted that whilst the amplitude and noise properties of
this starfield have been designed to
loosely approximate real stellar ellipticity data, no attempt has been
made to ensure that the spatial variation in the underlying
$e_{\mt}$ resembles coherent physical patterns as caused by telescope focusing
errors, coma or other optical phenomena (see, e.g.,
\citealp{jarvisetal08}).  This is essentially a separate, although
important, issue: in the following investigation the aim is merely to
test whether $D_1(r)$ and $D_2(r)$ help in the correct identification
of an appropriate fitting scheme $f_{\mm}$ for the simulated
starfield, i.e.\ whether the most successful fit is based upon a fifth
order bivariate polynomial, matching the level of input spatial
structure.  

Nonetheless, the analysis still has practical relevance as
it must be hoped that typical PSF patterns mostly fall within the
space of possible starfields in the random prescription described
above (otherwise polynomial fitting itself is likely to fail).  In
Sections \ref{sect:monte} \& \ref{sect:arbitrary} the testing will be extended to large sample
of random starfields, precisely in order to explore the success of
$D_1(r)$ and $D_2(r)$ for a variety of input `true' signals.  The
remainder of this Section will be concerned with the success of these
new diagnostics for the starfield of Figure \ref{fig:whisker}.

\subsection{Fits to the starfield ellipticities}\label{sect:starffits}

The ellipticity variation in the simulated starfield was then least-squares fit
using each of a set of four different bivariate polynomial fitting
schemes $f_{\mm}$: third, fourth, fifth (matching the input polynomial
order) and sixth order simple polynomials, performing a fit to each component
of ellipticity independently. The use of simple rather
than Chebyshev polynomials for fitting is immaterial, as each can be formally
expressed in terms of the other (at equivalent order) via exact linear
transformations in the polynomial coefficients. 
The best-fitting model was found in
each case using Singular Value Decomposition (SVD) as implemented by
the IDL routine \textsc{svdfit.pro}, itself based upon the
\textsc{svdcmp.f} routine within \emph{Numerical Recipes}
\citep{pressetal92}.  The SVD fitting process has the property of
minimizing numerical round-off error and matrix singularity problems
when attempting to solve underdetermined systems of equations, and
thus even in the case of an overfitting $f_{\mm}$ will produce the best possible solution.  This is important as the
diagnostic tests presented should be as insensitive as possible to
numerical issues.

So as to provide a clear example of overfitting, the starfield
was artificially split into eight regions associated with hypothetical
CCD chips.  These `chip regions' can be seen in Figure
\ref{fig:whisker}, bordered by dashed lines.  This is done to
illustrate the increased potential for overfitting when modelling
`chipwise', as is commonly done in weak lensing (see, e.g,
\citealp{fuetal08}, who independently model the PSF anisotropy in each
of the 36 CCD chips in the 1 deg$^2$ CFHTLS-W field of view). However,
such work only uses low-order polynomials for each chip, and it is
also interesting to observe the behaviour of $D_1(r)$ and $D_2(r)$ in
cases of severe (unrealistic) overfitting.  Fitting chipwise we also
begin to explore the question of whether modelling
schemes that cannot perfectly fit $e_{\mt}$, i.e.\ a `wrong' model, 
may be validated as practically sufficient given limited data.
For this chipwise fit we adopt schemes $f_{\mm}$ that use first, second, third and
sixth order bivariate polynomial surfaces, each being fit to each chip
independently.

\subsection{Comparison with simple modelling diagnostics}\label{sect:starfieldtests}
\begin{figure*}
\begin{center}
\psfig{figure=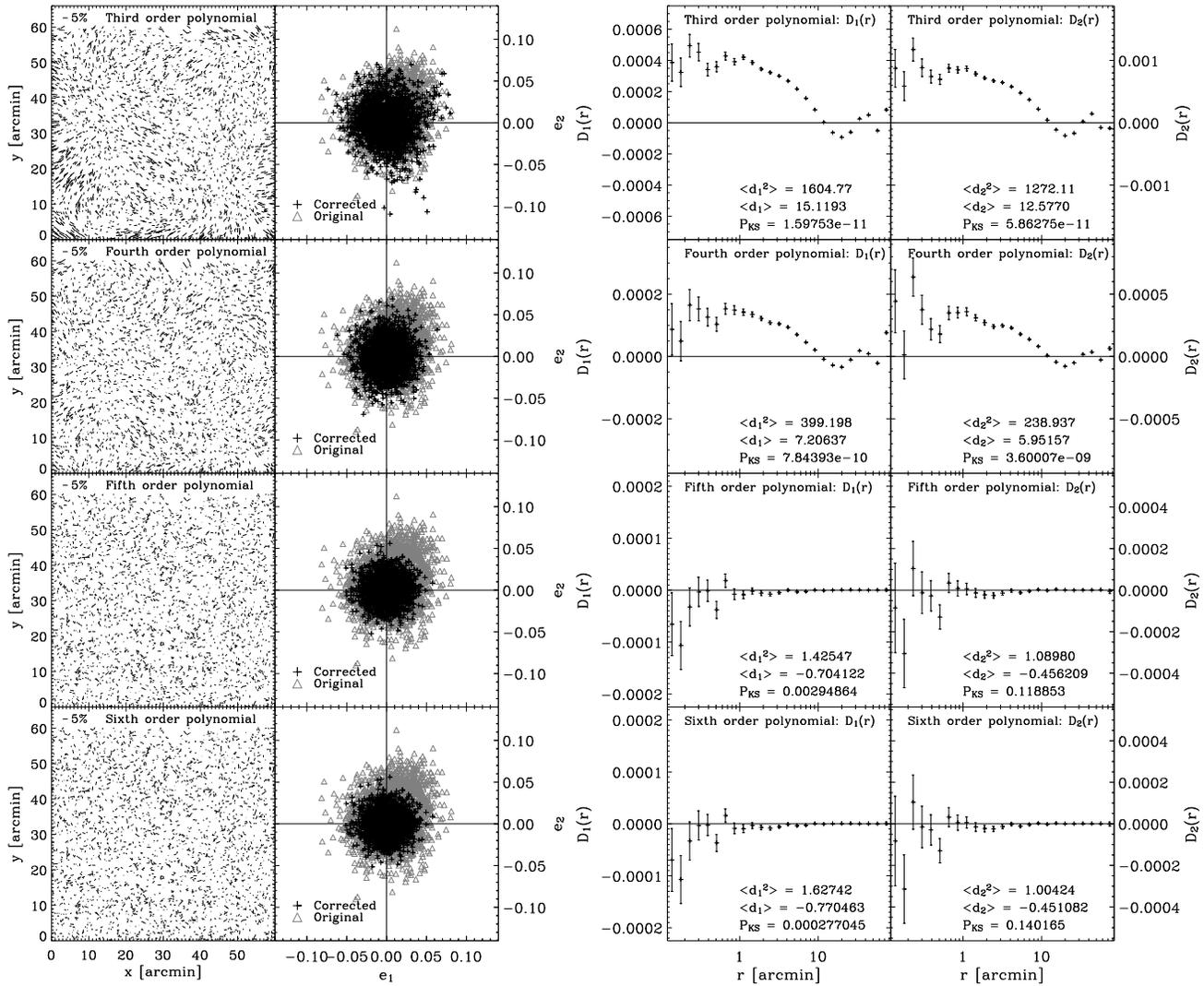,width=17.4cm,angle=0}
\caption{\label{fig:starfdiag} Traditional weak lensing tools for
  displaying the starfield PSF model fit results (left panels) and the
  corresponding $D_1(r)$ and $D_2(r)$ diagnostic correlation functions
  (right panels). As described in Section \ref{sect:starf}, the input
  PSF anisotropy $e_{\mt}$ was a fifth order bivariate Chebyshev
  polynomial surface in
  each component of ellipticity.}
\end{center}
\end{figure*}

\begin{figure*}
\begin{center}
\psfig{figure=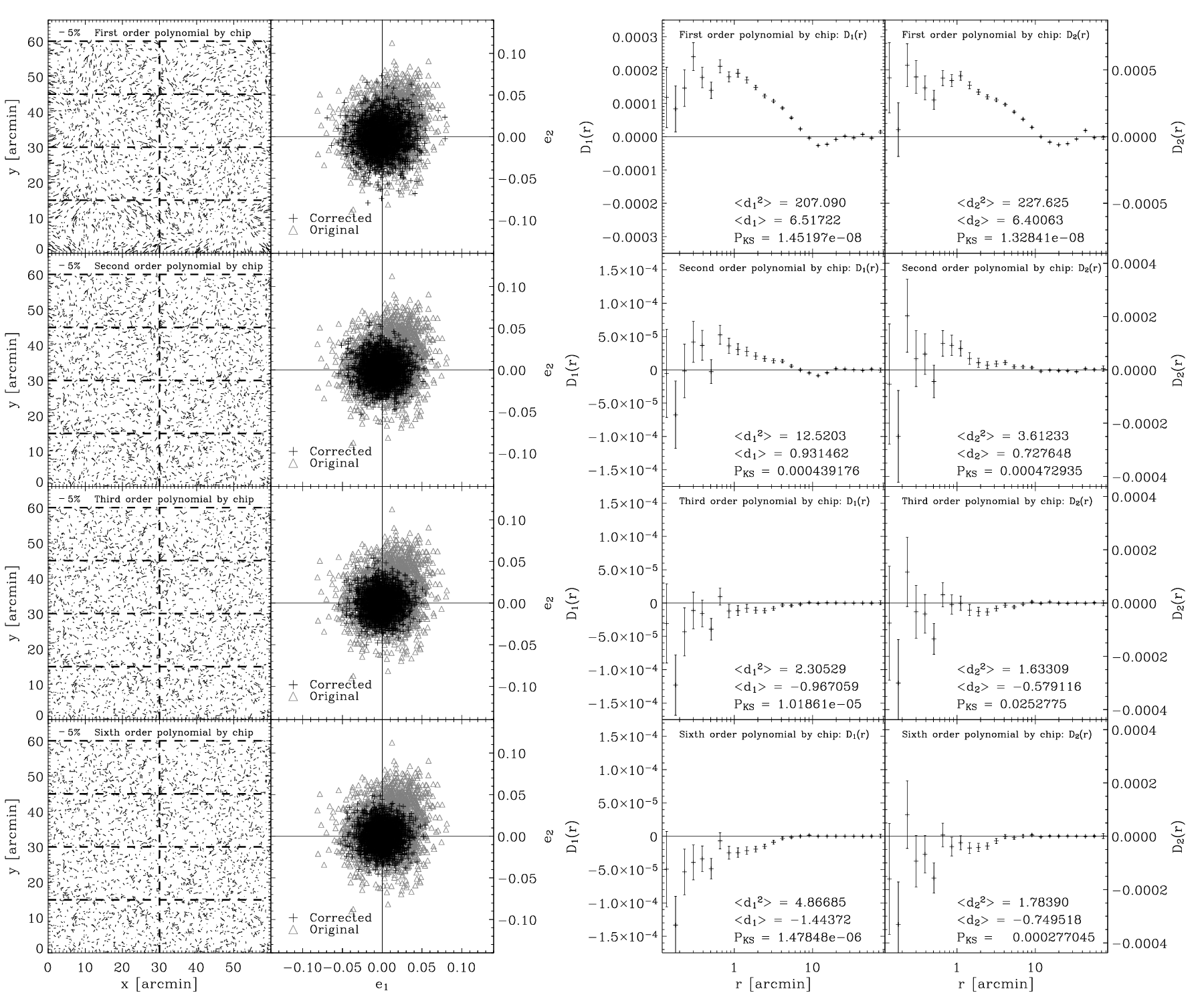,width=17.4cm,angle=0}
\end{center}
\caption{\label{fig:starfdiagchip} Traditional weak lensing tools for
  displaying the starfield PSF model fit results (left panels) and the
  corresponding $D_1(r)$ and $D_2(r)$ diagnostic correlation functions
  (right panels).  These results are for the independent chipwise fits
  to each of the eight chip regions in the starfield.  As described in
  Section \ref{sect:starf}, the input PSF anisotropy $e_{\mt}$ was a
  fifth order bivariate Chebyshev polynomial surface 
in each component of ellipticity.}
\end{figure*}
Having fit the 2500 simulated starfield measurements of $e$ using a
set of models $e_{\mm}(\xb ; f_{\mm})$, for a variety of different
fitting schemes $f_{\mm}$, the diagnostic functions $D_1(r)$ and
$D_2(r)$ are calculated using the formula in equation
\eqref{eq:xipprac}. Measurements are divided into 25
logarithmically-spaced angular bins between 7 arcsec and 1.4 deg.
This binning scheme was chosen as fairly representing both small and
large scale information. By varying these values 
it was also verified that $D_1(r)$ and
$D_2(r)$ were stable in regards to this choice, which was found to be
true once sufficient numbers of bins were used as to be able to
explore small scale correlations. Uncertainties were then calculated
as the standard error upon the mean value from all the pairs within
each bin, therefore \emph{not} taking the correlation in values
between neighbouring bins into account.  Calculations of the $D_1(r)$
and $D_2(r)$ diagnostics, for each scheme $f_{\mm}$, can be seen in
the right hand panels of Figure \ref{fig:starfdiag} for the global fit
and Figure \ref{fig:starfdiagchip} for the chipwise fit.

Whilst it should be noted that residual-residual correlations have already been
used by some groups to rule out or justify PSF models (see, e.g.,
\citealp{hoekstra04,vanwaerbekeetal05,hoekstraetal06,schrabbacketal07}),
and that these will be compared to $D_1(r)$ and $D_2(r)$ in the following Section,
the left hand panels of Figures \ref{fig:starfdiag} \&
\ref{fig:starfdiagchip} show comparative examples of more simple diagnostics used in the
 past to display the results of PSF model fits (see, e.g.,
\citealp{hoekstraetal98,heymansetal05,schrabbacketal07}).
The far left hand panels of Figures \ref{fig:starfdiag} \&
\ref{fig:starfdiagchip} show `whisker plots' of $e - e_{\mm}$
(referred to as corrected ellipticities) to depict the random nature
of fitting residuals. As can be seen by comparison with the
corresponding $D_1(r)$ plots, correlation and anticorrelation may
exist that is difficult to accurately quantify by eye, although
qualitatively there are traces of correlation in the third and fourth
order fit whisker plots.  A correlation
analysis of some sort is nonetheless clearly desirable, just as has been argued
previously \citep{hoekstra04}. 
The inner left hand
panels depict the distribution of `original' ellipticities $e$ in
comparison to the distribution of the corrected ellipticities
$e-e_{\mm}$; once again these plots are more difficult to
quantifiably interpret than $D_1(r)$ and $D_2(r)$, which show
more markedly different behaviour with $f_{\mm}$.  The
question remains as to whether these new diagnostics are behaving as predicted in Section \ref{sec:theory}.

Examining first the global fit results shown in Figure
\ref{fig:starfdiag} it can be seen that both $D_1(r)$ and $D_2(r)$
show clear residual positive and negative correlations for the
underfitting models, as predicted.  Both diagnostics appear
to be broadly consistent with zero for the fifth and sixth order fits.  
For the chipwise
fits of Figure \ref{fig:starfdiagchip} the results are similar but
show an interesting extra feature.  
While both diagnostics rule out a first
order chipwise fit as clearly underfitting, and the second order more
marginally, the third order chipwise fit shows slight
evidence of anticorrelation on small scales.  This is then seen more clearly
when the fit is taken to sixth order chipwise, particularly for the
$D_1(r)$ diagnostic.  These results suggest an overfit to the data for
third and sixth order chipwise fits, at least
according the reasoning presented in Section
\ref{sec:over}: this is a reasonable conclusion given knowledge of the number of
degrees of freedom in the initial model, and 
could not have been so easily diagnosed using current methods.

Furthermore, the fact that neither the second nor third order chipwise 
fits show perfect consistency with zero suggests that \emph{none} of the
schemes chosen in this artificial chipwise splitting of the field is
best suited to modelling the data, also a reasonable conclusion.  However, it may 
be that this apparent inconsistency is instead caused by chance and the fact that $\expect{m^*m}(\rb)$ is no longer isotropic, 
since the artificial chips are rectangular.  Nonetheless, even the
possibility that $D_1(r)$ and $D_2(r)$ might allow general modelling schemes to be
iteratively improved by correcting flaws such as the wrong choice of 
fitting function family, or the unnecessary splitting into independent chips, is of practical interest when fitting to an $e_{\mt}$ of unknown functional form.  Fitting to an arbitrary underlying field using a `wrong' (or, more accurately, incomplete) basis will be explored further in Section \ref{sect:arbitrary}, in which
polynomial fits will be made to randomly generated fields with only an average
power spectrum specified.

In summary, for the simple example presented in Figure
\ref{fig:starfdiag}, it appears that 
the degree of agreement with $D_1(r) = D_2(r) =
0$ is a potentially useful aid to model selection when compared with
simple diagnostic tools that have been used in the past.
In Figure \ref{fig:starfdiagchip} both diagnostics help rule out models
that would clearly be under- or overfitting, but there is no
scheme that performs perfectly once the field is artificially split
into chips. This fact gives hope that correlation analyses of this
sort might guide, in a directed manner, iterative improvements to 
modelling where there is no clear physical motivation for selecting a
given scheme (in the case presented it might motivate the decision not
to model chipwise, for example).
 It is now instructive to compare the results of this
analysis with the more complex diagnostics of PSF modelling that have
been discussed in the literature, which are also based upon correlations in
residuals, to see what may be added by the approach presented here.

\subsection{Comparison with the aperture mass dispersion and other
  related correlation diagnostics}

The use of residual correlation diagnostics in a similar form to $D_1(r)$
and $D_2(r)$ is not new,
having been advocated by \citet{hoekstra04} in the form of the
aperture mass dispersion estimator $\mathcal{M}(r)$ 
defined in equation (16) of \citet{schneideretal02p1} (see also
\citealp{crittendenetal02}; \citealp*{schneideretal02bmode}), a filtered
combination of $\xi_+(r)$ and $\xi_-(r)$ that gives an unbiased
estimator of the cosmological aperture mass dispersion:
\begin{equation}
\expect{M^2_{\textrm{ap}}}(r) = \frac{1}{2}  \! \int \! \frac{r' \dif
  r'}{r^2} \! \!
\left[ \xi_+(r') T_+\left( \frac{r'}{r^2} \right) + 
\xi_-(r') T_-\left( \frac{r'}{r} \right) \right]
\end{equation}
and
\begin{equation}
\expect{M^2_{\perp}}(r) = \frac{1}{2} \! \int \! \frac{r' \dif
  r'}{r^2} \! \!
\left[ \xi_+(r') T_+\left( \frac{r'}{r^2} \right) - 
\xi_-(r') T_-\left( \frac{r'}{r} \right) \right],
\end{equation}
where the functions $T_{\pm}(x)$ are non-zero only for $x < 2$ and
are given in \citet*{schneideretal02bmode}.
The aperture mass dispersion has
the useful property that it allows a decomposition into `E' and `B'
mode contributions (which are $\expect{M^2_{\textrm{ap}}}(r)$ and
$\expect{M^2_{\perp}}(r)$ respectively) 
to the correlation signal, the former of which only 
is produced by a simple scalar mass potential (although B-modes can be
created by contamination to the shear correction, intrinsic alignments
of source galaxies etc., see \citealp{schneider06}).  This measure was 
then employed by \citet{vanwaerbekeetal05} and \citet{hoekstraetal06}
to find optimal schemes for modelling the spatial variation of the PSF
anisotropy, aiming towards $\mathcal{M}(r) =0$ for the residuals
between modelled and measured PSF anisotropies.  In both studies the
chosen PSF fitting scheme was that which minimized the aperture mass
dispersion of its residuals.

\begin{figure}
\begin{center}
\psfig{figure=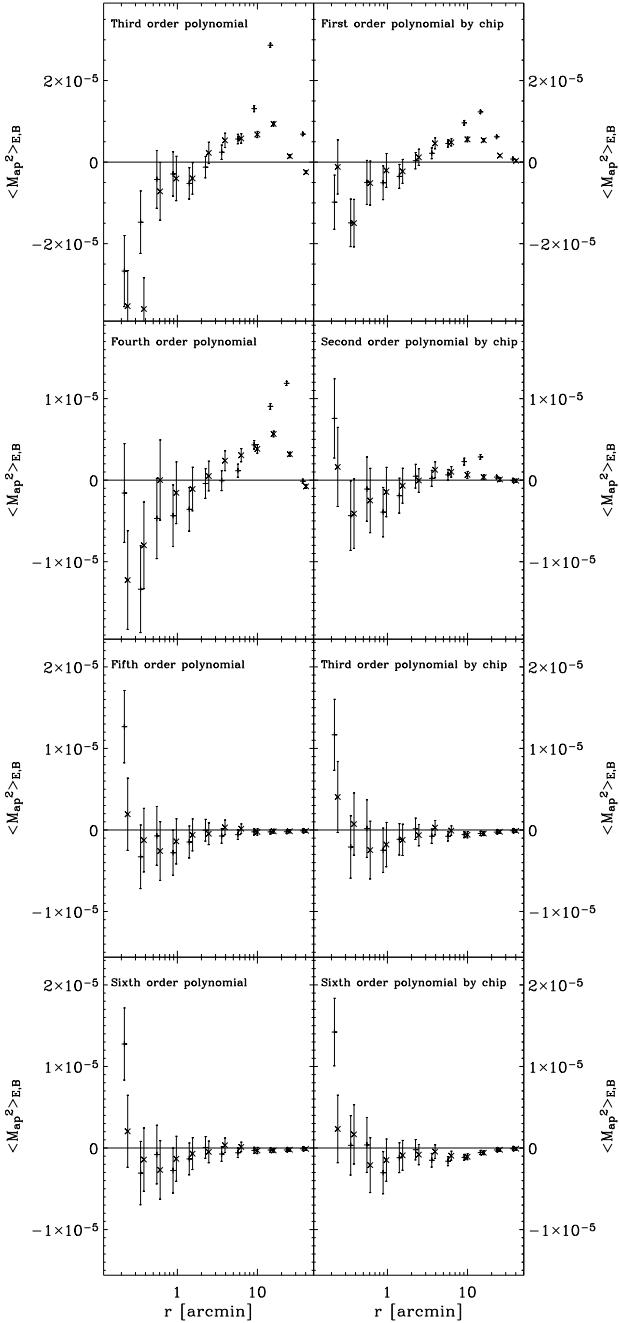,width=8.4cm,angle=0}
\caption{\label{fig:map2} Aperture mass dispersion statistic for
the residuals of the fits in Sections \ref{sect:starffits}
\& \ref{sect:starfieldtests}, corresponding to the diagnostics
plotted in Figures \ref{fig:starfdiag} \& \ref{fig:starfdiagchip}. The
crosses resembling addition signs represent the E mode signal and those
resembling multiplication signs the B mode signal, which are offset
slightly along the abscissa for clarity. Left hand side panels are for
the simple polynomial fits to the whole field, whereas the right hand
side panels show the dispersion in the chipwise fit residuals.}
\end{center}
\end{figure}
In Figure \ref{fig:map2} the aperture mass dispersion is plotted for
the residuals of the fits described in Sections \ref{sect:starffits}
\& \ref{sect:starfieldtests}, corresponding to the diagnostics
plotted in Figures \ref{fig:starfdiag} \& \ref{fig:starfdiagchip}. 
The results are similar to those of $D_1(r)$ and $D_2(r)$, in that
there are clear correlations for the underfitting cases.
The signal to noise is reduced, a known feature of this statistic
(e.g., \citealp{hoekstraetal06,schneider06,fuetal08}),
and a means of positively identifying overfitting versus underfitting is less
clear.  This is because the aperture mass 
filtering of $\xi_+(r)$ and $\xi_-(r)$, which provides a desirably
clean statistic when considering lensing due to physical mass
distributions, makes the interpretation of correlated modelling residuals (via
simple arguments such as those presented in Section \ref{sec:theory})
more complicated. One can say that correlation or
anticorrelation exists, but saying which and why is less clear.

There is a suggestion in the bottom right hand side panel of Figure
\ref{fig:map2}, for the sixth order chipwise fit, that the negative
values of $\expect{M^2_{\textrm{ap}}}_{E/B}$ are an indication of the
overfitting model in the same way as the negative values of $D_1(r)$
and $D_2(r)$ at small scales; although with apparently less overall signal than
$D_1(r)$.  Justifying this hypothesis 
using extensions to the arguments of Section
\ref{sec:theory} may be possible, and would immediately
identify a clearly overfitting PSF model in the study performed by
\citet{hoekstraetal06} (Figure 7 of that work): a not impossible
finding given their independent fitting
of second order polynomials to each of 36 chips across the CFHTLS
Megacam field. Whether this sheds more insight
than $D_1(r)$ and $D_2(r)$ upon the PSF model quality is unclear, however, and
even unlikely given the lower signal to noise of the aperture mass
measures (although see \citealp{fukilbinger09}).

Nevertheless, as $\expect{M^2_{\textrm{ap}}}$ (or the more recently-derived ring statistics $\expect{\mathcal{R} \mathcal{R}}_{E/B}$, see \citealp*{schneiderkilbinger07,eifleretal09,fukilbinger09}) provides a model-independent method for E/B mode decomposition, it is in many ways a preferred choice for constraining cosmological parameters. Therefore, it will undoubtedly be of use to express residual correlations in the PSF model in terms of such quantities to quantify the contamination to the measured shear signal (e.g., \citealp{hoekstra04}; see also Section \ref{sect:relation}). But in the diagnosis of poor modelling, and in the directed manner in which simplifications or increases in complexity to the PSF model can be motivated, the interpretational clarity of $D_1(r)$ and $D_2(r)$ is an advance upon the use of the aperture mass dispersion.

\citet{schrabbacketal07} also present a residual correlation
analysis as a diagnostic of PSF modelling. These authors demonstrate 
that their chosen PSF model minimized 
the two functions $|\xi_+(r) + \xi_-(r)|$ and
$|\xi_+(r) - \xi_-(r)|$ when measured upon model-data stellar
ellipticity residuals, after randomly drawing stars from dense stellar
fields (estimating the impact upon cosmic shear 
by scaling to an equivalent shear correction using randomly-drawn
$P^g$ values from a galaxy population; see \citealp{schrabbacketal07}).
Overfitting was not a concern for these authors, who built a family of
robust and detailed \emph{HST} PSF models using dense stellar fields 
and then selected
from this family via a maximum-likelihood match to the far fewer 
stars available in the galaxy survey images.  As such, no attempt was made to
interpret the sign of the resulting functions and the absolute values
alone sufficed as a diagnostic. Again, the interpretability of $D_1(r)$ and
$D_2(r)$ in terms of over- and underfitting, via the arguments in
Section \ref{sec:theory}, is of significant additional value.

\subsection{Quantifying the agreement with $D_1(r) = D_2(r)=0$}\label{sect:quantify}

These results are interesting, but in order to test their repeatability it will be necessary to find some method of compressing the data and quantifying the qualitative visual appraisal so far conducted.  Ideally there would be a means of ascribing a single, easily-calculated number to the $D_1(r)$ and $D_2(r)$ results, describing how well a given model matches the desired $D_i(r)=0$ behaviour.  An obvious example is a chi-squared-like measure, but a true chi-squared is impossible without knowledge of the covariance between bins of $D_i(r)$, and these covariance matrices may in general only be estimated \emph{post hoc} (often imperfectly) via a statistical jackknife or bootstrap. The calculation of even a single $D_1(r)$ or $D_2(r)$ takes some short time, varying with the square of the number of data points, and so these processor-intensive bootstrap techniques quickly become prohibitively expensive.  In this Section we examine practical possibilities for a cheaper alternative to a full chi-squared measure of the $D_i(r)=0$ hypothesis.

To assign simple numbers to these results, in effect
searching for a better proxy to the `appraisal-by-eye' 
performed in the previous Section, the following 
quantities are defined:
\begin{eqnarray}
(d_i)_j & = &\frac{D_i(r_j)}{\sigma_{D_i}(r_j)} \label{eq:dji}, \\ 
\expect{d_i} & = & \frac{1}{N_{\textrm{bins}}} \sum_j (d_i)_j \label{eq:dj}, \\
\expect{d^2_i} & = & \frac{1}{N_{\textrm{bins}}} \sum_j (d_i)^2_j \label{eq:dj2}.
\end{eqnarray}
Here $i=1,2$
so as to specify the $D_1(r)$ or $D_2(r)$ diagnostic respectively,
$j$ denotes each discrete bin of angular scale $r_j$, and the error
estimate on each $D_i(r_j)$ is denoted as $\sigma_{D_i}(r_j)$.
As discussed above, it should be noted that the values for
neighbouring bins of $D_i(r)$ are correlated and the uncertainties
$\sigma_{D_i}(r_j)$ do not take this into account: it would thus be
dangerous to  associate equation \eqref{eq:dj2} with any sort 
of true chi-squared measure of the goodness of fit to a desired $D_i=
0$ scenario. 


However, use may be made of the fact that the correlation between bins for
the diagnostics must be expected to be minimized for those successful
models that approach $D_i=0$: this can be simply argued by considering that
$m$ is expected to become approximately random in these cases.  Also
of use is the fact that the most important task at hand is merely to
select a best-fitting model using a single number that quantifies this
success.  It might still be hoped that \eqref{eq:dj} and
\eqref{eq:dj2} are useful as a way of \emph{ranking} competing models.
The overall normalization of $\expect{d_i}$ and $\expect{d^2_i}$ for
failing cases would, when ranking, be less important than the relative
normalization as compared to successful cases.  

Another means of quantifying the agreement with $D_i(r)=0$ is via the
the Kolmogorov-Smirnov (KS) test (see, e.g., \citealp{pressetal92,lupton93}),
suggested to the grateful author by the anonymous referee.  In
the null hypothesis of a well-fitting model we may approximate that
the individual values of $(d_i)_j$ are each independently described by a Gaussian distribution with zero mean and unit variance. 
The maximal difference between the
empirical cumulative distribution function (derived from the data) and
the cumulative distribution function of the null hypothesis can then be
used to assess the goodness of fit. The probability $P_{\textrm{KS}}(d_i)$ that the
null hypothesis could produce at least the maximal difference seen may then be
calculated from a series approximation to the Kolmogorov distribution (although note that the use of this distribution is not strictly accurate, given that the `data' $(d_i)_j$ we use are derived quantities, and so this measure should be rightly interpreted only as an approximate guide; see \citealp{pressetal92}).

The values of each of the $\expect{d_i}$, $\expect{d^2_i}$ and $P_{\textrm{KS}}(d_i)$ statistics, for each fit to the
starfield, are given in Figures \ref{fig:starfdiag} \&
\ref{fig:starfdiagchip}.  It would be naively expected that the best
fitting $f_{\mm}$ would show minimum values of $\expect{d^2_i}$
and $|\expect{d_i}|$, and a maximum $P_{\textrm{KS}}(d_i)$: this is seen in the case
of $D_1(r)$, but the $D_2(r)$ results marginally favour the
overfitting sixth order polynomial in this case.  
All statistics strongly rule out lower order fits. 
For the chipwise fits the situation is more complex, due to their
being no fitting model which shows clear consistency with $D_i(r)=0$.  
Second and third order fits are variously preferred. As
discussed in Section \ref{sect:starfieldtests}, this is perhaps due to
there being no $f_{\mm}$ in this artificially split case that can
reproduce all features of the ellipticity field without some overfitting 
redundancy.

All three statistics show some promise in being able to
approximately describe the extent to which $D_i(r)=0$.  More sophisticated approaches could certainly be explored, particularly
given the information regarding the expected signatures of under- and
overfitting described in Sections \ref{sec:under} \& \ref{sec:over}. 
This extra information might be perhaps be used
within a Bayesian framework to select models, particularly if the
modeller wished to rule out either an under- or overfitting model at
any cost. This may be fertile ground for future work, as it is clear that
the three statistics $\expect{d^2_i}$, $|\expect{d_i}|$, and $P_{\textrm{KS}}(d_i)$
were chosen above all for their immediate simplicity and clarity.

These results lead to the possibility of asking another question: can model
selection via the statistics described here be successfully
repeated for a variety of input $e_{\mt}$ fields?  
A single example field is hardly strong
evidence for the utility of $D_1(r)$ and $D_{2}(r)$ 
in more general cases, so many more fields
must be tested. The broad success of the three simplifying
statistics described in this Section
offers hope that the author might be spared from the need to
visually inspect many plots of $D_1(r)$ and $D_2(r)$, and avoid the inevitable subjectivity in such an approach.  Moreover,
the necessity and difficulty 
of performing a full and stable statistical jackknife
calculation (in order to estimate covariances) may 
be avoided.  The following Section tests the
success of $\expect{d_i^2}$, $|\expect{d_i}|$ and $P_{\textrm{KS}}(d_i)$ as criteria for
correctly identifying appropriate modelling order using a large number 
of randomly generated $e_{\mt}$ anisotropy fields.

\section{Monte Carlo testing using a suite of simulated anisotropy maps}\label{sect:monte}
The repeatability the results of Section \ref{sect:testcase}, which
hinted at the utility of $D_i=0$ as a model selection criterion, will
now be tested using a large number of simulated $e$ fields.  To
test the model selection for input physical signals of varying spatial
complexity, each component of the input $e_{\mt}$ is modelled as an
$n$th order bivariate Chebyshev polynomial surface 
as described by equation \eqref{eq:ai},
where the upper limit of the double sum is now $j + k \le n = 1, 2, \ldots
,7$.   For each $n$ a
suite of $N_{\mMC}=500$ randomly generated starfields is then created
using a procedure directly analogous to that described in Section
\ref{sect:starf}.  Then, for each
of the 500 simulated starfields of a given input order $n$, a set of
best-fitting models for the anisotropy map are constructed as
described in Section \ref{sect:starffits}; this time, however, a
greater range of bivariate polynomial fitting schemes $f_{\mm}$ are
explored, ranging in fitting order from $m = 1, \ldots, 7$.

The same process is also performed using the artificial chipwise division of the field of view when fitting.  Finally, the resulting $D_1(r)$ and $D_2(r)$ statistics were calculated as described in Section \ref{sect:starfieldtests} for each input order $n$ and fitting order $m$, for each of the 500 random starfields, and the $\expect{d^2_i}_{\mmin}$, $|\expect{d_i}|_{\mmin}$ and $P_{\textrm{KS}}^{\mmax}(d_i)$ criteria discussed above were then used to select the most appropriate modelling order.

\subsection{Results}
\begin{figure}
\begin{center}
\psfig{figure=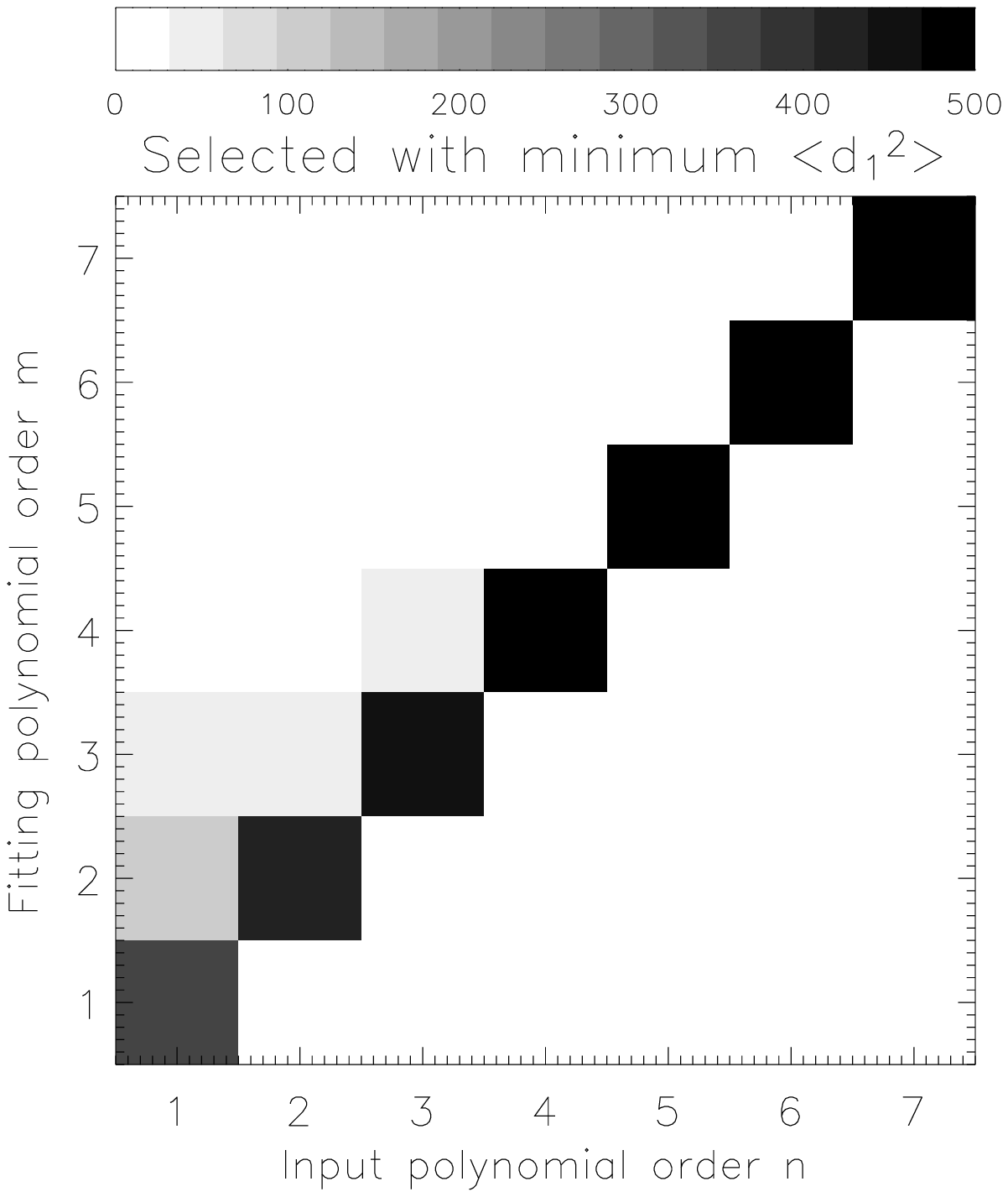,width=4.15cm,angle=0}
\psfig{figure=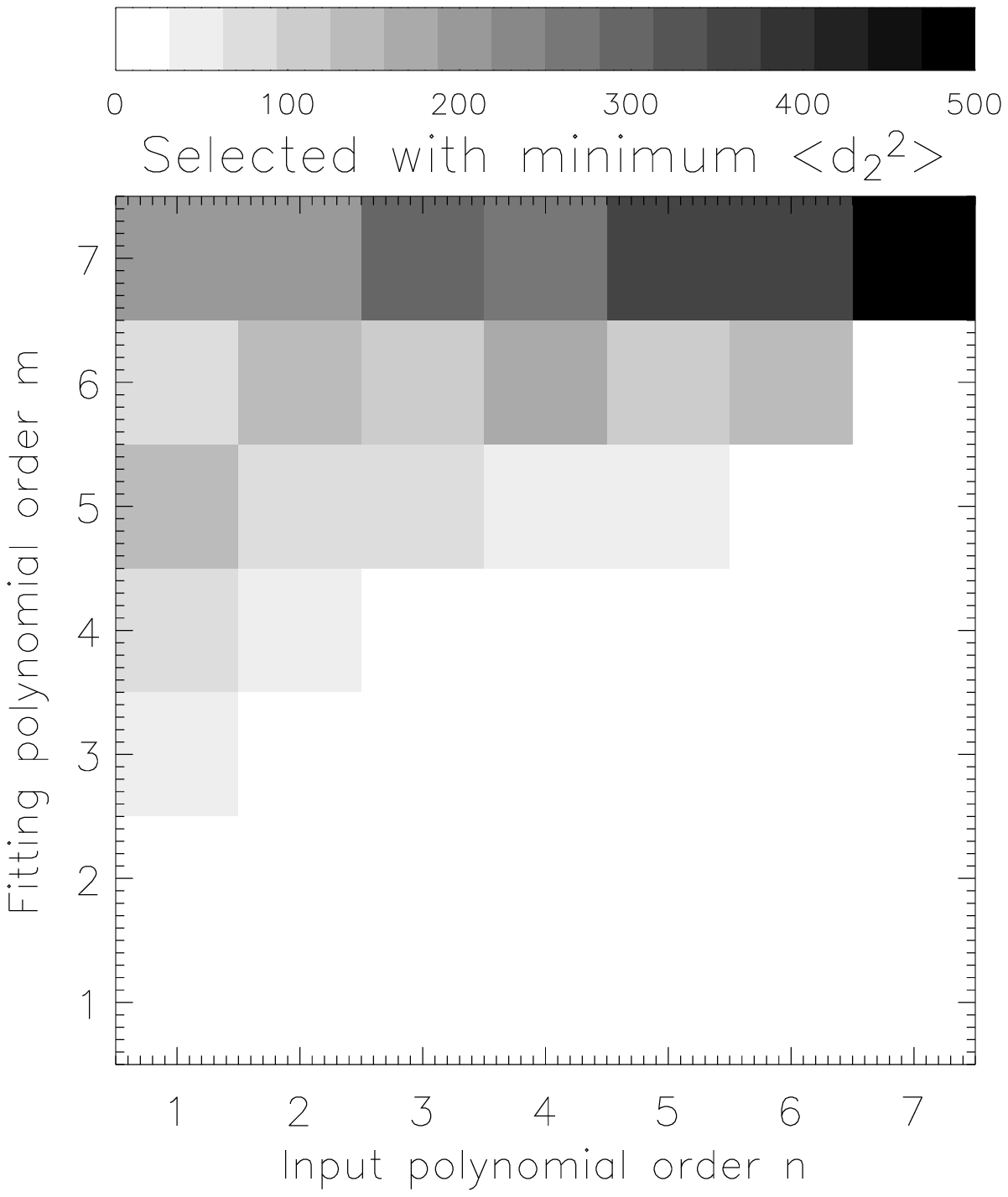,width=4.15cm,angle=0}
\psfig{figure=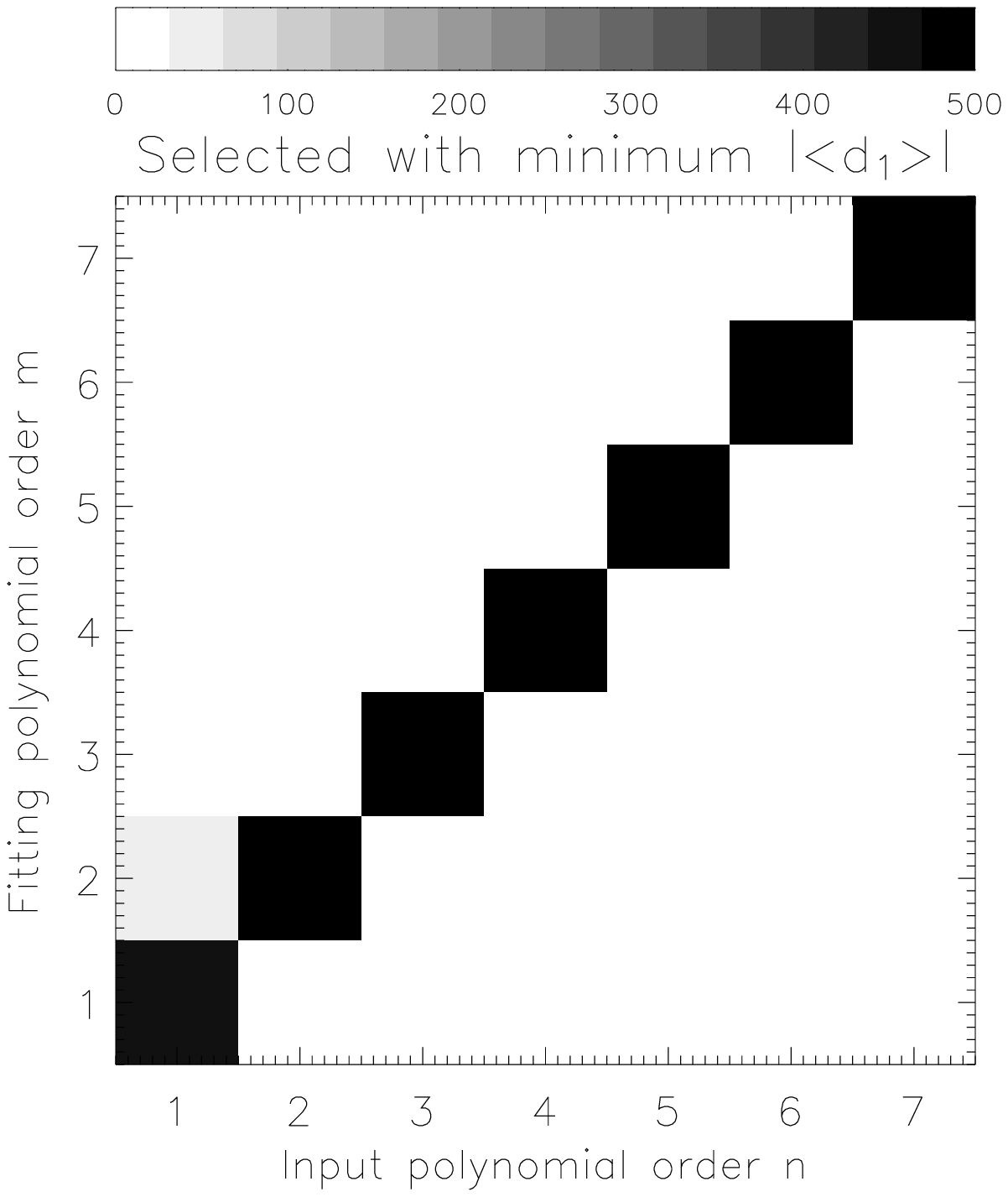,width=4.15cm,angle=0}
\psfig{figure=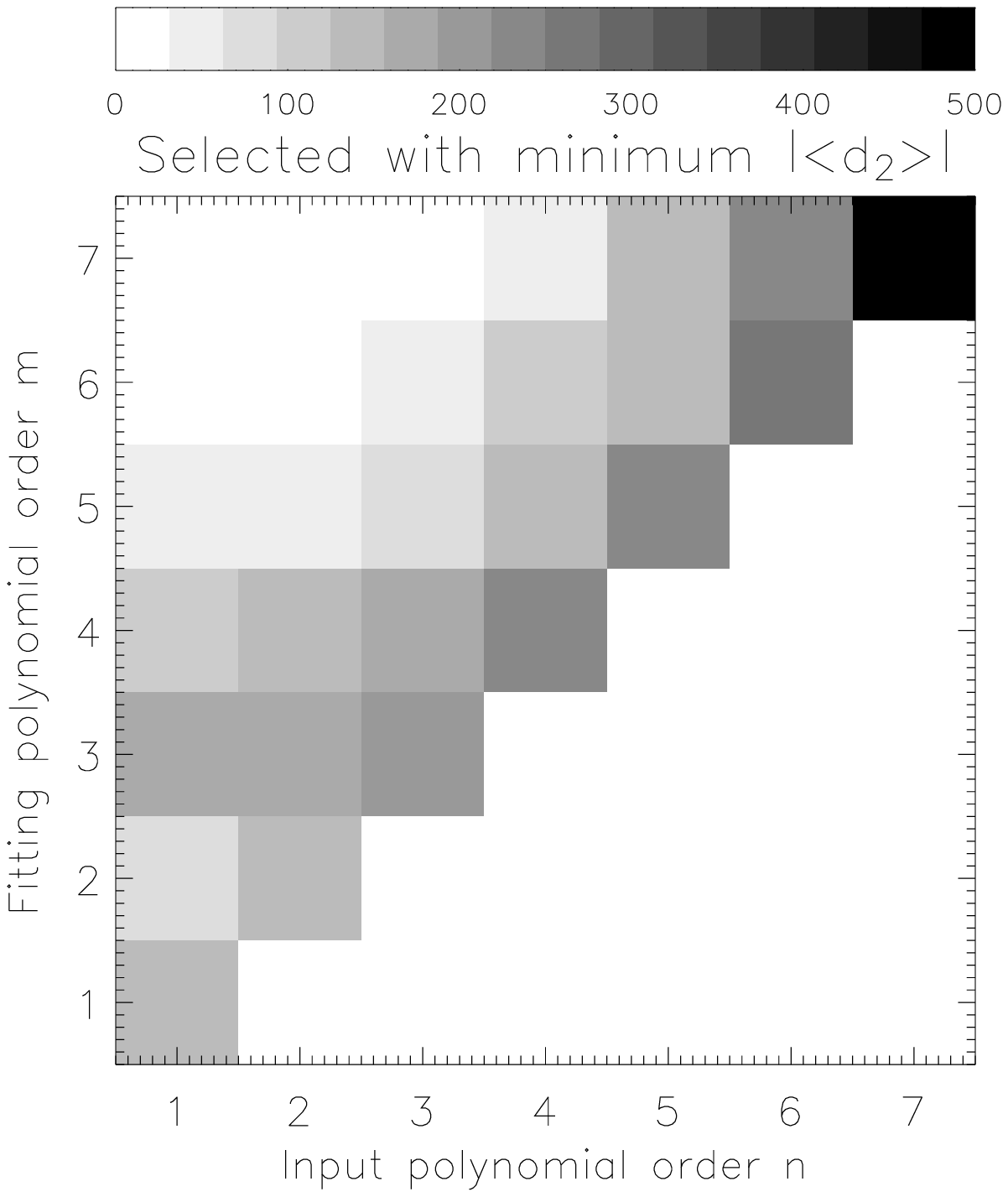,width=4.15cm,angle=0}
\psfig{figure=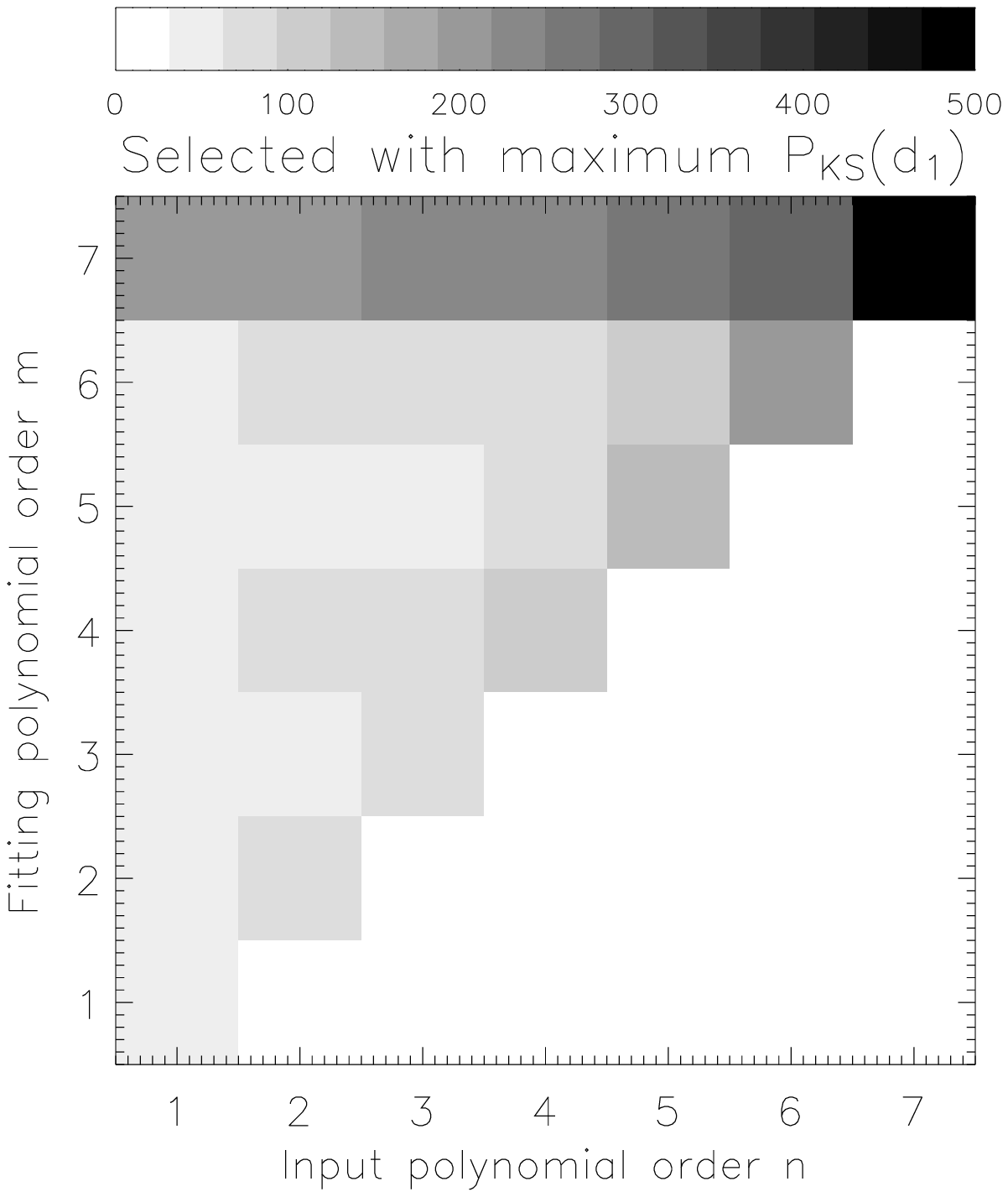,width=4.15cm,angle=0}
\psfig{figure=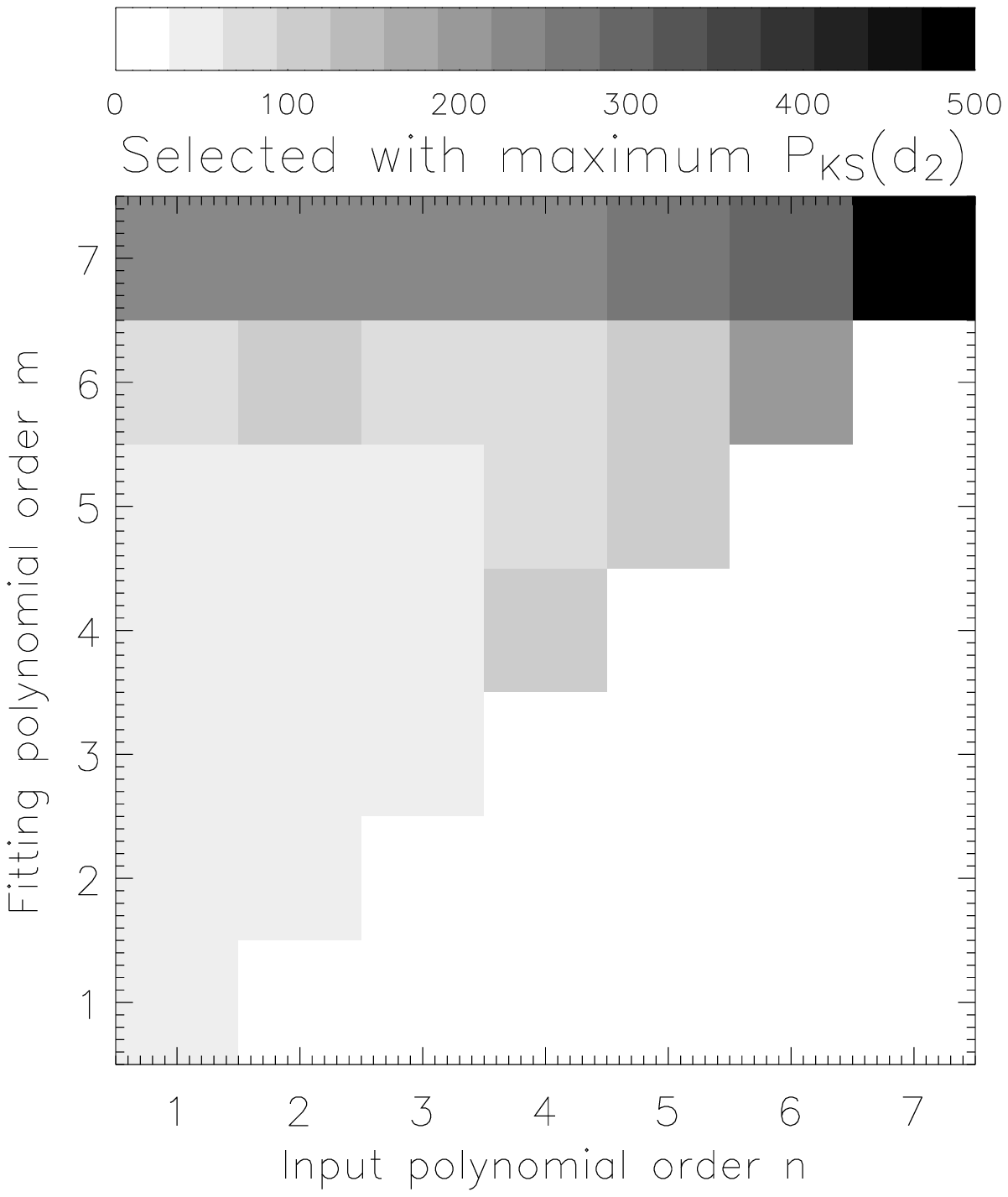,width=4.15cm,angle=0}
\caption{\label{fig:dmin} Number of model schemes selected as best
  fitting using the $\expect{d^2_1}_{\mmin}$ (upper left),
  $\expect{d^2_2}_{\mmin}$ (upper right), $|\expect{d_1}|_{\mmin}$
  (mid left), $|\expect{d_2}|_{\mmin}$ (mid right),
  $P_{\textrm{KS}}^{\mmax}(d_1)$ (lower left) and $P_{\textrm{KS}}^{\mmax}(d_2)$ (lower
  right) criteria, as
  a function of input polynomial order $n$. The total number of fields
  simulated at each input order was $N_{\mMC} = 500$. Figures
  \ref{fig:dmin} \& \ref{fig:dminchip} were
  created using a modified version of the {\sc
    shapelets\_plot\_image.pro} routine from the publicly-available
  \emph{Shapelets} software \citep{masseyrefregier05}. }
\end{center}
\end{figure}
\begin{figure}
\begin{center}
\psfig{figure=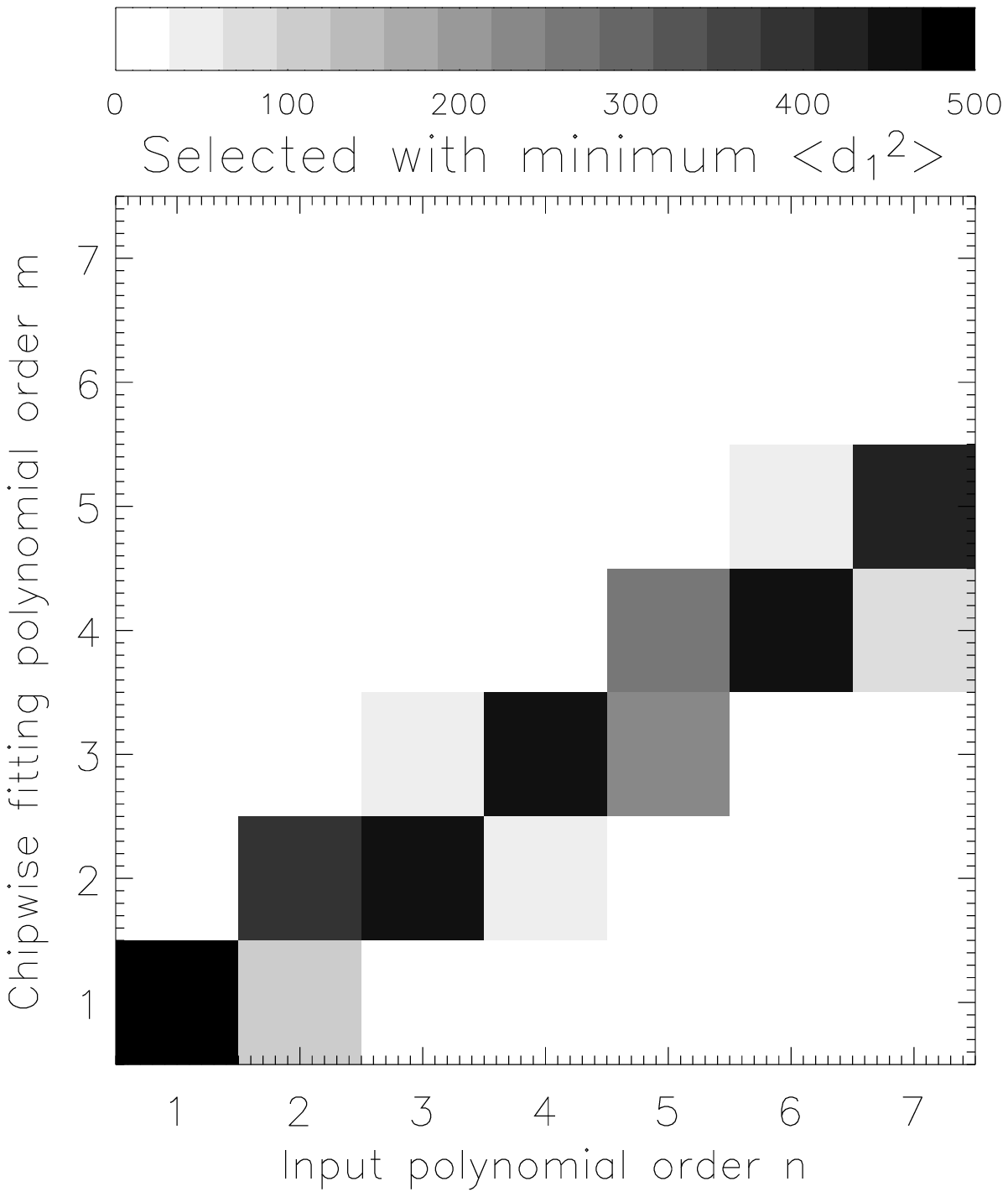,width=4.15cm,angle=0}
\psfig{figure=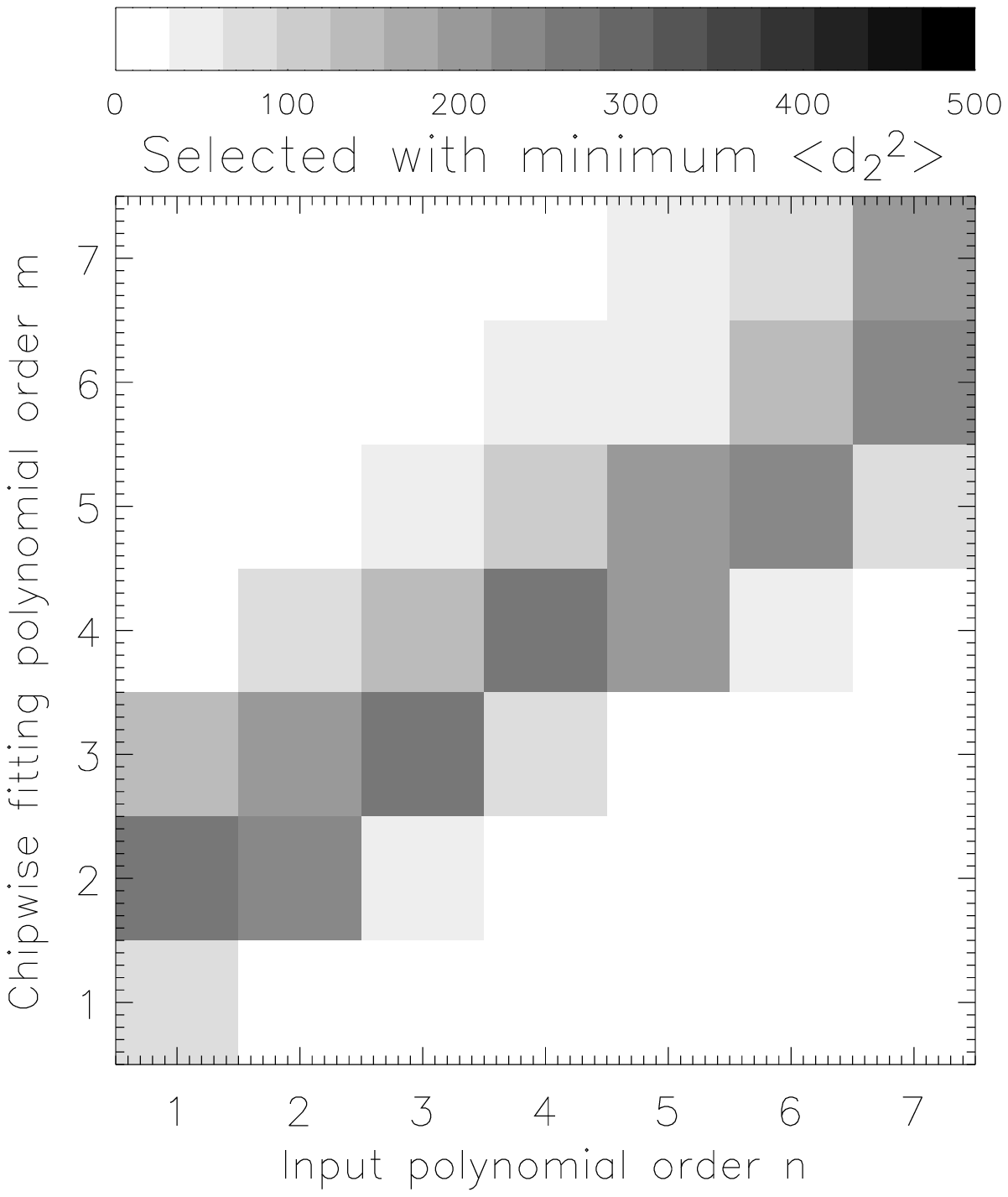,width=4.15cm,angle=0}
\psfig{figure=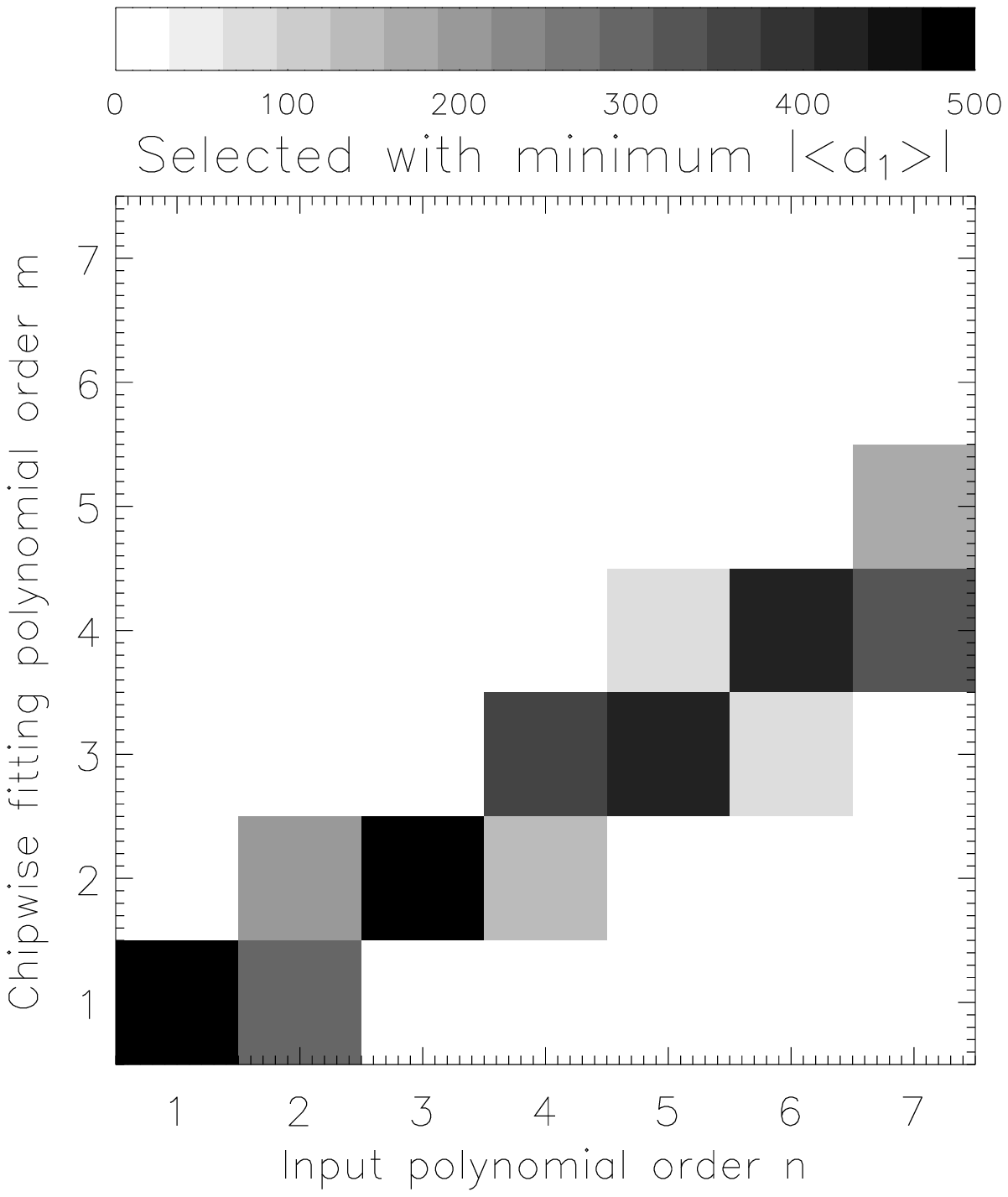,width=4.15cm,angle=0}
\psfig{figure=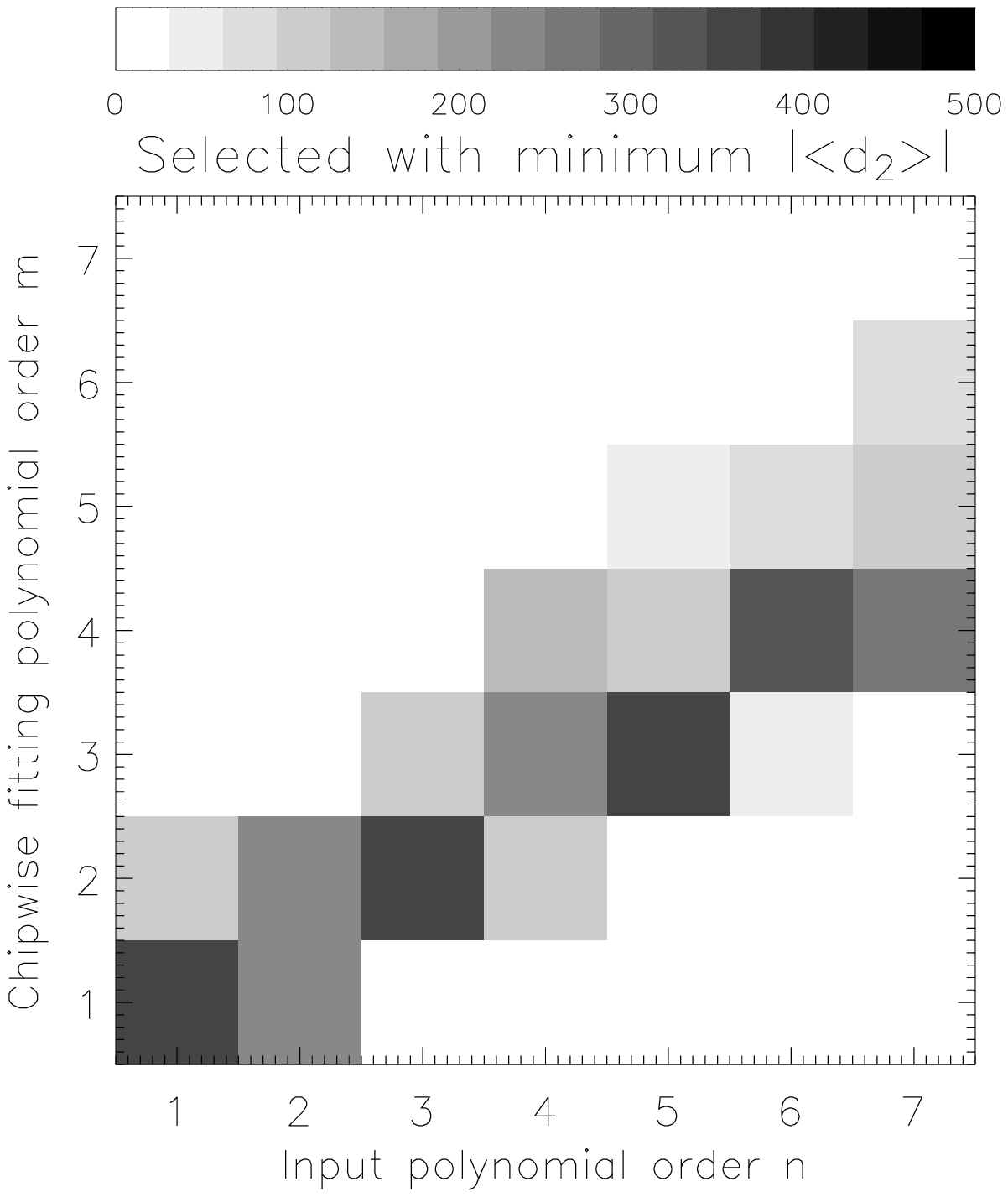,width=4.15cm,angle=0}
\psfig{figure=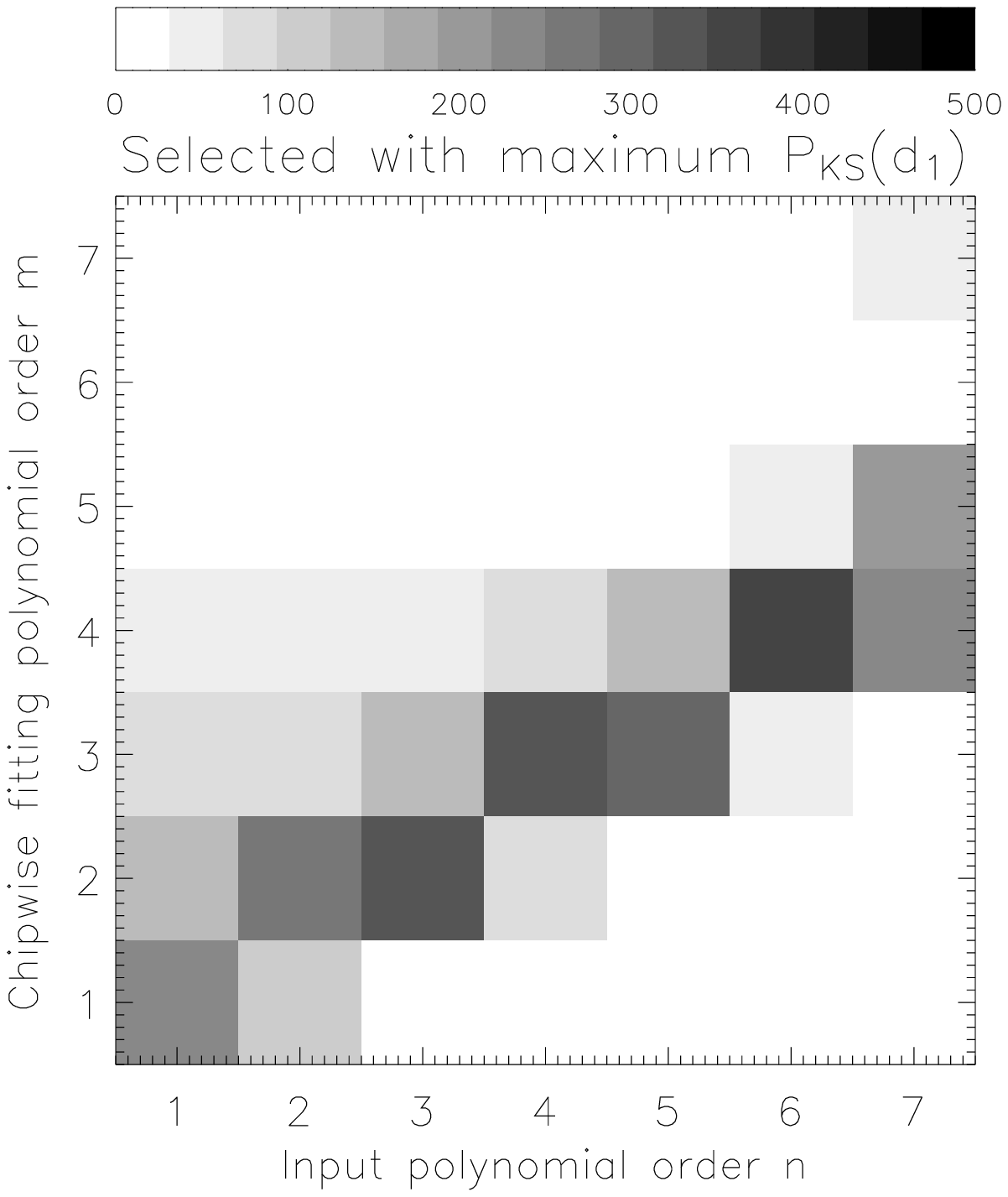,width=4.15cm,angle=0}
\psfig{figure=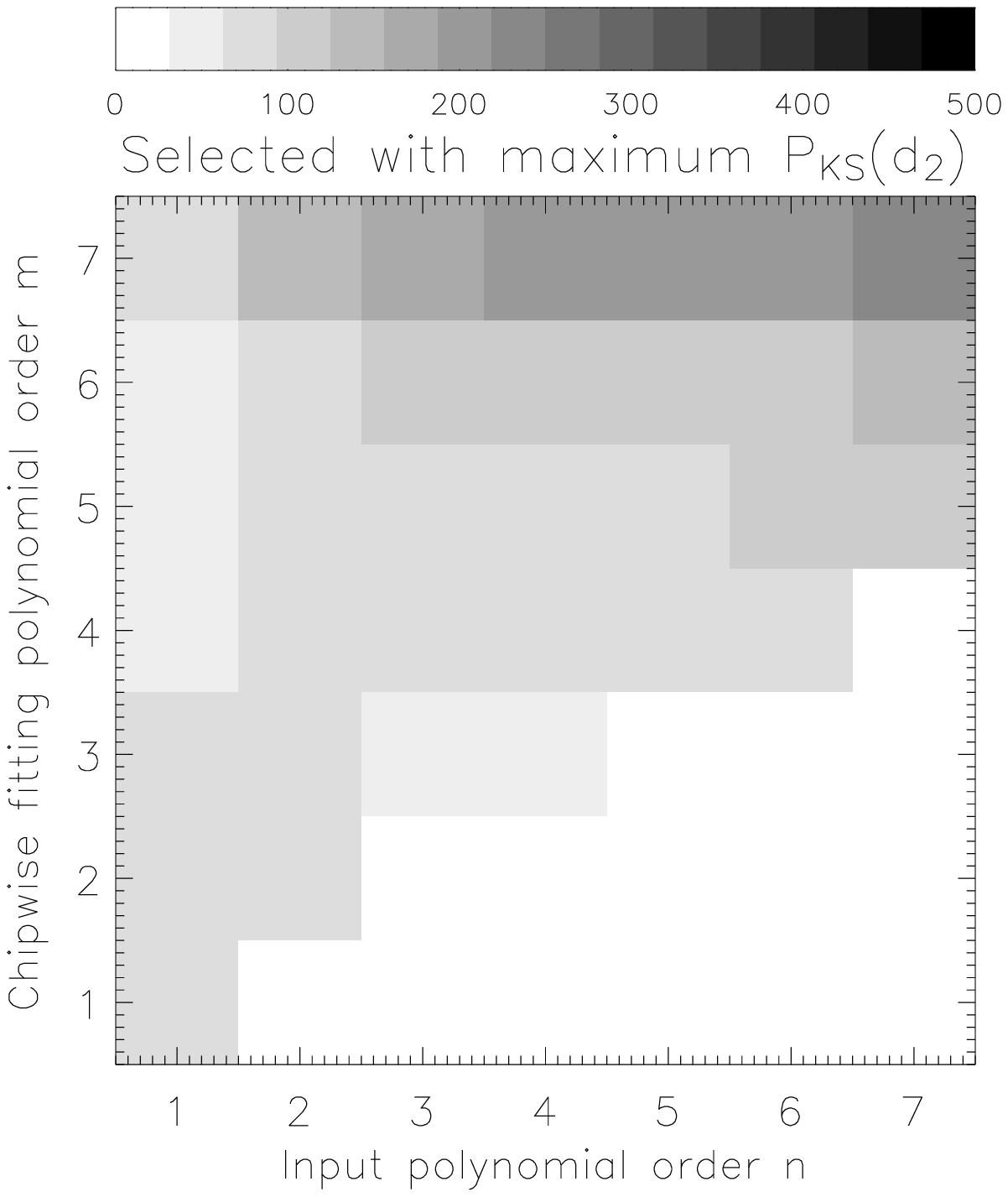,width=4.15cm,angle=0}
\caption{\label{fig:dminchip} 
Number of chipwise model schemes selected as best
  fitting using the $\expect{d^2_1}_{\mmin}$ (upper left),
  $\expect{d^2_2}_{\mmin}$ (upper right), $|\expect{d_1}|_{\mmin}$
  (mid left), $|\expect{d_2}|_{\mmin}$ (mid right),
  $P_{\textrm{KS}}^{\mmax}(d_1)$ (lower left) and $P_{\textrm{KS}}^{\mmax}(d_2)$ (lower
  right) criteria, as
  a function of input polynomial order $n$.
The total number of fields
  simulated at each input order was $N_{\mMC} = 500$.}
\end{center}
\end{figure}
The number of times a global fitting scheme of order $m$ was chosen as best
representing the simulated starfield of input order $n$ is shown in
Figure \ref{fig:dmin}, quantifying the success of the
$\expect{d^2_i}_{\mmin}$, $|\expect{d_i}|_{\mmin}$ and
$P_{\textrm{KS}}^{\mmax}(d_i)$ criteria for
this simple Monte Carlo test.  The same results are shown for the
chipwise fits in Figure \ref{fig:dminchip}.

For the range of input models and statistics tested it
appears that $D_1(r)$ is typically a cleaner and more reliable model selection
diagnostic than $D_2(r)$, which in both the global and chipwise fits
is noisier and appears to show a
biased relative preference for overfitting models in most
cases. $\expect{d^2_1}_{\mmin}$ and $|\expect{d_1}|_{\mmin}$ show
striking success in selecting the correct fitting order in Figure
\ref{fig:dmin}, but not so their $D_2(r)$ counterparts.  The performance
of $P^{\mmax}_{\textrm{KS}}(d_i)$ is more disappointing overall and seems
insensitive to identifying overfits when fitting globally, although 
in Figure \ref{fig:dminchip}
there is evidence of it correctly ruling out the more extreme chipwise
overfits, at least for $D_1(r)$. $|\expect{d_1}|_{\mmin}$ appears to show moderate preference for lower order fits than $\expect{d^2_1}_{\mmin}$, particularly at lower orders, perhaps due to the close cancelling of correlation and anticorrelation on different scales (c.f.\ Section \ref{sec:under}). 
 Nonetheless, all statistics appear to rule out underfitting models successfully.

The chipwise results show better rejection of overfitting models, and therefore greater overall agreement between the results for different statistics.  This is not surprising when one considers that for chipwise fits the number of degrees of freedom in the fitting model increases more rapidly with increasing order $m$ than in the global case.  The comparison of the chipwise and global results therefore suggests that, for those statistics which failed to rule out overfits, the problem was more of relative sensitivity than a total failure.  The $P_{KS}^{\mmax}(d_i)$ statistic performs the worst, and \citet{pressetal92} suggest a possible reason for this: the sensitivity of the KS test of deviations from the c.d.f.$(d_i)$ is not independent of $d_i$, and is in fact most sensitive around the median value.  This makes the test least sensitive in the wings of the probability distribution, and hence to the outliers which may give the clearest indication a poor fit. This is one plausible explanation, and \citet{pressetal92} describe possible alternatives to the KS test that attempt to mitigate this problem.  Another reason might be the incorrect use of the Kolmogorov distribution in the test, since $(d_i)_j$ are quantities derived from the data rather than direct data themselves (see Section \ref{sect:quantify}; also \citealp{pressetal92}).

These results leave open an avenue for potentially valuable further study, as an optimal way for identifying $D_1(r) = D_2(r) = 0$ is clearly not yet found. Despite this fact, the results of Figures \ref{fig:dmin} \& \ref{fig:dminchip} are extremely encouraging, for $\expect{d^2_1}_{\mmin}$ and $|\expect{d_1}|_{\mmin}$ in particular.  The results $D_2(r)$ were generally less successful, and seemed to show an insensitivity to overfitting in particular.  Possible reasons for this are now briefly discussed.



\subsection{Understanding the relative failure of $D_2(r)$}\label{sect:d2fail}
In the Monte Carlo tests performed it appears that the selection
criteria based upon $D_2(r)$ show some shortcomings that need to be
explored. Firstly, they often prefer overfitting
models when compared to the $D_1(r)$ results (biased when compared to the truth for
Figure \ref{fig:dmin}). Secondly, both $\expect{d^2_2}$ and
$|\expect{d_2}|$ are noisy.  This latter can perhaps be partially
explained by the fact that $D_2(r)$ is itself often more noisy than
$D_1(r)$, a natural consequence of the value and variation of $e$ being typically
larger than that of $e-e_{\mm}$ (see equations \ref{eq:d1} and \ref{eq:d2});
the biased results warrant further investigation.

Complete answers as to why $D_2(r)$ performed relatively poorly in
this simple experiment almost certainly lie outside the capability of
this paper, as this behaviour may depend non-trivially upon many
factors.  Possible contributions could be: the number of stars
simulated per field; the overall signal to noise for the $e$ field;
the geometry of the chosen field of view; the nature of the typical
pattern described by the simulated $e_{\mt}$ (the chosen polynomials,
even when randomly generated, do not come close to being able to
describe arbitrary surfaces); the degree of variance and covariance
within and between bins of $D_2(r)$. Whilst all these factors may
potentially be limiting in the correct circumstance, the last is
worthy of some further discussion in particular as it is one aspect in
which $D_2(r)$ may differ significantly from $D_1(r)$.

Evidence of overfitting is generally seen at small scales, and it has
been seen that on large scales $D_i \simeq 0$ even for drastic
overfits (e.g.\ Figure \ref{fig:starfdiagchip}). At these scales the
quoted errors upon $D_2(r)$ will still be large when compared to those
for $D_1(r)$, because of the typically greater values of $e$ as
compared to $e-e_{\mm}$, meaning that contributions to
$\expect{d_2^2}$ on these scales are overly suppressed.  Moreover, the
use of $\expect{d_i^2}$ as a means of ranking competing models will
work best in situations where the covariance matrix is totally
diagonal, and will correspond in such cases to a chi-squared-like
measure of goodness of fit to $D_i=0$.  Conversely, strong
off-diagonal values might produce exactly the results seen in Figures
\ref{fig:dmin} \& \ref{fig:dminchip} for the $D_2(r)$ diagnostics. If
false, the assumption of weakly correlated uncertainties between bins,
which is implicit for the utility of $\expect{d^2_i}$ and
$|\expect{d_i}|$, will lend undue weight to the supposed success of
$D_i=0$ on large scales and therefore undue credence to the
overfitting model itself.  If the covariance matrix for $D_2(r)$
contains significant off-diagonal terms, even in the case of a
successful model, this may be cause for the systematic preference
towards overfitting models seen in Figures \ref{fig:dmin} \&
\ref{fig:dminchip}.

In that case, however, it must also be seen if there is a systematic
reason why the covariances of $D_1(r)$ and $D_2(r)$ differ in their
diagonality.  One important difference is apparent from the very
definitions given in equations \eqref{eq:d1} and \eqref{eq:d2}: the
first diagnostic is the residual-residual autocorrelation, whilst the
second is a measure of the data-residual cross-correlation.  It may be
that $D_2(r)$ will therefore suffer from strong covariances even for
cases approaching $D_2(r) = 0$, due to the physically correlated
nature of the data $e$, whereas such correlations in $D_1(r)$ will
inevitably decrease as the model tends towards $D_1(r) = 0$.  Another
difference may lie in the angle averaging, performed over all pairs
separated by distance $r$, implicit in the definition of the
correlation functions thus far used in this paper (see, e.g.,
equations \ref{eq:xipm} and \ref{eq:xipprac}).  It can be imagined
that $D_2(r)$, being more strongly dependent upon the non-isotropic
field $e_{\mt}$, might suffer from greater uncertainty and even bias
due to the angle averaging that is implicit in the way $D_1(r)$ and
$D_2(r)$ are calculated.

These are plausible explanations for the effects seen but,
unfortunately, this discussion will remain essentially un-concluded without further investigation.  Estimating \emph{post hoc} the
covariance matrices of correlation functions such as $D_2(r)$ is
expensive with current computing resources and, as the covariance of
$D_2(r)$ will depend non-trivially upon the starfield $e_{\mt}$, the
calculation would need to be made many times within the current
simulation framework.  Pessimistically, even with such work we may in fact be learning more about the properties of any simulated $e_{\mt}$ than about $D_i(r)$. Despite these issues regarding $D_2(r)$, it has certainly been demonstrated that both $D_i(r)$ show some potential for diagnosing modelling quality, and that $D_1(r)$ appears outstandingly successful in the tests so far conducted. A next step is to enlarge the space of potential $e_{\mt}$ fields to include more general, arbitrary structure, and to see how $D_1(r)$ and $D_2(r)$ perform when a perfect fit may no longer be possible.

\section{Fitting arbitrary spacial patterns}\label{sect:arbitrary}

So far in this paper the testing of the $D_1(r)$ and $D_2(r)$ diagnostics has been for cases in which a perfect fit to the data was possible at some modelling order. The chipwise fitting made the situation somewhat more interesting by balancing this against an increased propensity to overfit: for example, whilst it might be necessary to fit a fifth order polynomial chipwise to formally recover \emph{all} aspects of an underlying fifth order global $e_{\mt}$, it may not always be justified by the data available.  This was reflected in the results of Figure \ref{fig:dminchip}, which preferred lower order chipwise fits even though such models might not be able to perfectly reproduce all aspects of the input PSF variation.  In this Section we take the analysis one step further by introducing more arbitrary spacial patterns, to see whether the diagnostics can differentiate between more general variation in the degree of spatial structure.

Once again a Monte Carlo approach is taken and we build three suites of $N_{\mMC} = 500$ randomly generated starfields. For each field, each component of the underlying $e_{\mt}$ is modelled as a Gaussian random field with a specified average power spectrum.  These can be constructed by summing random phase Fourier modes across the 1 deg$^2$ field of view
\begin{equation}\label{eq:efourier}
(e_{\mt})_i = \sum_{n,m=1}^{N_{\textrm{limit}}} c_{inm} \cos({k_n x + \phi_n}) \cos(k_m y + \phi_m),
\end{equation}
where $k_n = n \pi$ deg$^{-1}$, and $i=1,2$ denotes the real and imaginary parts of $e_{\mt}$ respectively. The value $N_{\textrm{limit}} = 300$ was chosen to ensure that structure was fairly represented well below the typical separation for 2500 objects in a 1 deg $^2$ field (1/50 deg), although this slowed the field generation somewhat.  $\phi_n$ and $\phi_m$ are chosen to to be random uniform variables in the range [0, $2\pi$] and, having defined $k^2 = k_n^2 + k_m^2$, the coefficients $c_{inm}$ are then drawn from Gaussian random variables of zero mean and variances satisfying the constraint that the ensemble average power spectrum is given by a power law 
\begin{equation}
\expect{c_{inm} c_{in'm'}} = A k^{-\alpha} \delta_{nn'} \delta_{mm'}.
\end{equation}
For each of the three Monte Carlo starfield suites, different choices were made for the value of the power law slope $\alpha$: 11/3 and 11/6 (motivated by the Kolmogorov spectrum for atmospheric turbulence, e.g.\ \citealp{sasiela94}, relevant for ground-based lensing), and the intermediate value 11/4.  For each $\alpha$ the normalization $A$ was accordingly set so as to give equal power in the lowest order mode, specifically setting $\sqrt{\expect{c_{i11}^2}} = 0.01$.  This normalization, once stellar ellipticities were sampled at 2500 randomly-distributed  points and a Gaussian noise of $\sigma_N = 0.015$ added to each component (c.f.\ Section \ref{sect:starf}), once more recreated simulated starfields of realistic anisotropy for the CFHTLS-W.

\begin{figure}
\begin{center}
\psfig{figure=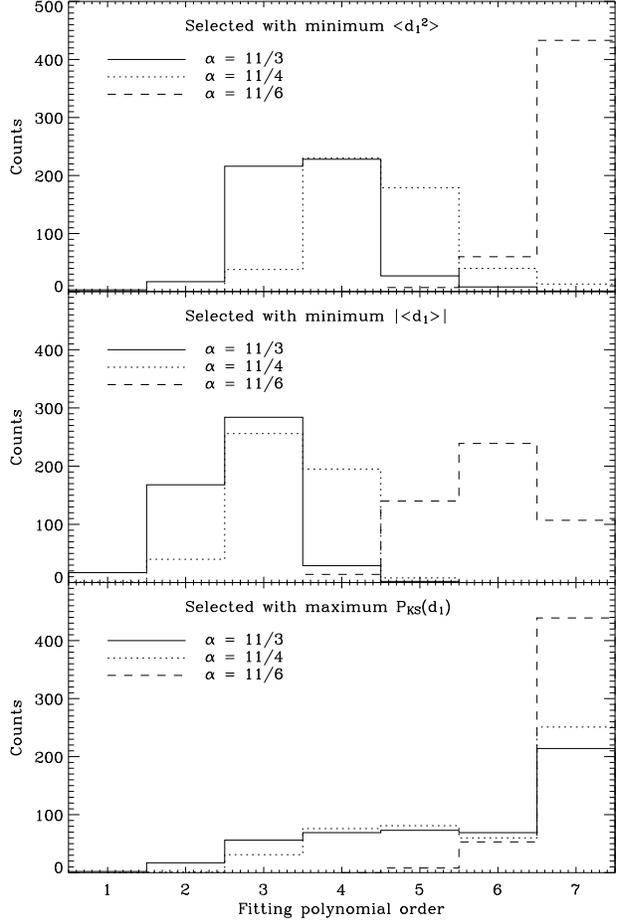,width=8.4cm,angle=0}
\caption{\label{fig:powerhistd1} Number of model schemes selected as best
  fitting using the $\expect{d^2_1}_{\mmin}$ (upper panel),
  $|\expect{d_1}|_{\mmin}$
  (middle panel), and $P_{\textrm{KS}}^{\mmax}(d_1)$ (lower panel) criteria, as
  a function of the input $e_{\mt}$ power spectrum slope $\alpha$. The total number of fields
  simulated for each input power spectrum was $N_{\mMC} = 500$.}
\end{center}
\end{figure}

\begin{figure}
\begin{center}
\psfig{figure=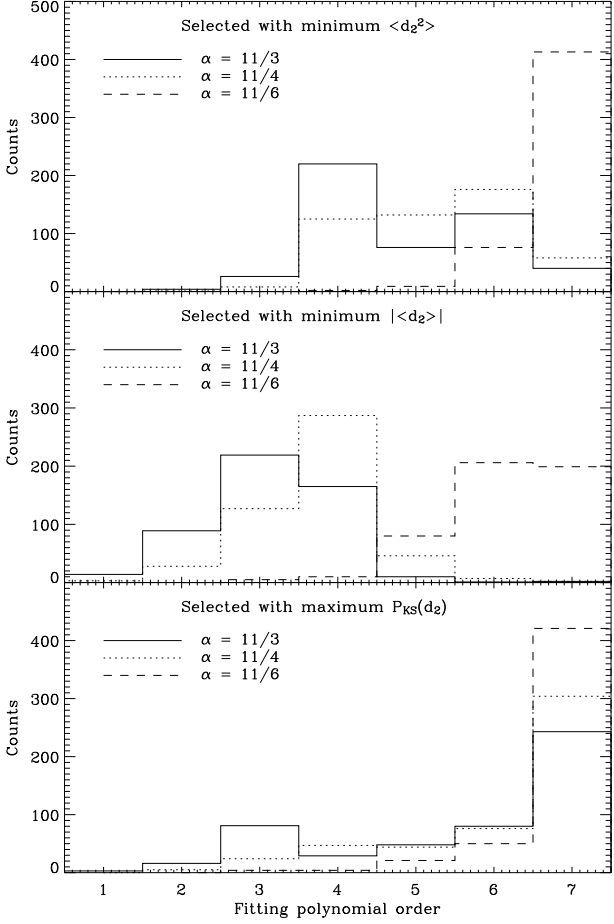,width=8.4cm,angle=0}
\caption{\label{fig:powerhistd2} Number of model schemes selected as best
  fitting using the $\expect{d^2_2}_{\mmin}$ (upper panel),
  $|\expect{d_2}|_{\mmin}$
  (middle panel), and $P_{\textrm{KS}}^{\mmax}(d_2)$ (lower panel) criteria, as
  a function of the input $e_{\mt}$ power spectrum slope $\alpha$. The total number of fields
  simulated for each input power spectrum was $N_{\mMC} = 500$.}
\end{center}
\end{figure}
As in Section \ref{sect:monte}, these simulation starfields were then fit using simple bivariate polynomials of order $1, \ldots, 7$, globally.
The $D_1(r)$ and
$D_2(r)$ statistics were calculated for each fitting order, and the
$\expect{d^2_i}_{\mmin}$, $|\expect{d_i}|_{\mmin}$ and $P_{\textrm{KS}}^{\mmax}(d_i)$ criteria used to select the most appropriate fitting model.  The results of these fits for $D_1(r)$ and $D_2(r)$ can be seen in Figures \ref{fig:powerhistd1} \& \ref{fig:powerhistd2}, respectively. 

Most importantly, there are clear differences in the distribution of preferred fitting order as a function of power spectrum slope: as would be hoped the shallowest $\alpha = 11/6$ case systematically prefers the highest order fits, suggesting even that higher than seventh order fits might be necessary in many of the starfields. The $\alpha = 11/4, 11/3$ cases peak lower (except for $P^{\mmax}_{KS}(d_i)$, which performs poorly again), with the steepest power spectrum preferring the lowest order fits.  The $\expect{d^2_i}_{\mmin}$ and $|\expect{d_i}|_{\mmin}$ measures all seem to be able to differentiate well between the degrees of spatial structure in these arbitrary underlying models. The tendency for $|\expect{d_i}|_{\mmin}$ to prefer lower order fits than $\expect{d_i^2}$ is seen again, as is the tendency for $D_2(r)$ to prefer higher order fits relative to $D_1(r)$.  It should be noted that this latter behaviour is strikingly modest when compared to the results of Section \ref{sect:monte}, perhaps due to a lesser degree of spatial correlation in the randomly-generated underlying $e_{\mt}$ (see discussion in Section \ref{sect:d2fail}).

The results of Figures \ref{fig:powerhistd1} \& \ref{fig:powerhistd2}, when taken alongside those for the polynomial starfields of Section \ref{sect:monte}, give compelling evidence for the utility of $D_i(r)$ as diagnostic tools for PSF modelling.  Here we have shown that they are able to discriminate between physical systems of varying complexity despite unknown or arbitrary underlying forms, and in Section \ref{sect:monte} it was shown that when the model approaches the `correct' form this too can be identified.  From simple arguments the diagnostics provide a way of making directed, iterative, and possibly automated improvements to both modelling accuracy and stability. Whilst the three descriptive statistics suggested in Section \ref{sect:quantify} remain imperfect in many aspects, there is scope for improving them.  All of this offers hope that $D_i(r)$ will be a useful tool for systematically improving PSF modelling in future weak lensing studies.  With this in mind, we now turn to a brief discussion of how these diagnostics may be related to requirements for cosmic shear measurement accuracy.

\section{Relation to cosmic shear predictions}\label{sect:relation}
The results presented so far in this paper compare various fitting models, but no attempt has been made to propagate these results through to the impact upon cosmic shear measurements.  In this Section we describe relevant results in the literature, and show how these allow values of $D_1(r)$ in particular to be related to the impact upon cosmic shear systematics.  This will be necessary to help define the limits and requirements for the quality of PSF modelling for future surveys.  $D_2(r)$ is less useful in this respect as it cross-correlates the modelling residuals with the telescope PSF $e$ variation, a filtering that on average will be unmatched to any cosmic signal.

In what follows significant use is made of results from \citet{paulinetal08}, adapted so to propagate the correlated effects of poor spatial modelling of the PSF variation into an impact upon measurements of cosmic shear.  These authors use a simple description based upon unweighted image moments to argue that, to first order in the quantities concerned, the systematic error $\delta e^{\textrm{sys}}$ upon a measured galaxy ellipticity will be given by
\begin{equation}
\delta e^{\textrm{sys}} \simeq \left(e_{\textrm{gal}} - e_{\textrm{PSF}}\right) \frac{\delta R^2_{\textrm{PSF}}}{R^2_{\textrm{gal}}} - \left( \frac{R_{\textrm{PSF}}}{R_{\textrm{gal}}} \right)^2 \delta e_{\textrm{PSF}}
\end{equation}
where $e_{\textrm{gal}}$ and $e_{\textrm{PSF}}$ are the ellipticity of the galaxy and the PSF in this region of sky, $R_{\textrm{gal}}$ and $R_{\textrm{PSF}}$ are the respective angular radii. $\delta R^2_{\textrm{PSF}}$ and $\delta e_{\textrm{PSF}}$ represent uncertainties in these aspects of the PSF due to poor or unstable modelling. Changing to the notation of Section \ref{sec:theory}, $\delta e_{\textrm{PSF}}$ can be immediately identified with the modelling inaccuracy $m$.  Considering only the second term (as the $D_i$ diagnostics cannot directly shed light on $\delta R^2_{\textrm{PSF}}$) we may write the associated error in the measured shear due to the poorly modelled PSF anisotropy as
\begin{equation}\label{eq:dgamma}
\delta \gamma^{\textrm{sys}} \simeq - \frac{m}{P^{\gamma}} \left(\frac{R_{\textrm{PSF}}}{R_{\textrm{gal}}} \right)^2,
\end{equation}
having defined the shear susceptibility factor $P^{\gamma}$ as in \citet{paulinetal08} (these authors take a value of $P^{\gamma} \approx 1.84$ as typical for a distribution of measured galaxy ellipticities).  
The measured shear for a region of sky may be then written as $\gamma = \gamma_{\mt} + \delta \gamma^{\textrm{sys}} + \delta \gamma^{\textrm{N}}$, where $\gamma_{\mt}$ is the `true' shear and $\delta \gamma^{\textrm{N}}$ is a random, assumed-unbiased noise term.
Using the expression \eqref{eq:xipm} to define a shear correlation $\xi^{\gamma}_+$ and taking equation \eqref{eq:dgamma} together with the inequality \eqref{eq:d1dodgy}, we may approximate the systematic impact due to a poorly modelled PSF anisotropy as
\begin{eqnarray}
|\delta \xi^{\gamma}_+|^{\textrm{sys}}(r) & \equiv & |\expect{\gamma^* \gamma}(r) - \expect{\gamma_{\mt}^* \gamma_{\mt}}(r)| \\
& = & \left| \frac{\expect{m^* m}(r)}{(P^{\gamma})^{2}} \expect{\left(\frac{R_{\textrm{PSF}}}{R_{\textrm{gal}}} \right)^4}  \right| \\
& \le &  \left| \frac{D_1(r)}{(P^{\gamma})^{2}} \expect{\left(\frac{R_{\textrm{PSF}}}{R_{\textrm{gal}}} \right)^4 } \right| \label{eq:d1sys1}.
\end{eqnarray}
Following the arguments of Section \ref{sec:over}, the last expression will tend towards an identity for underfitting models but place an upper limit on the systematic contribution from $m$ in the overfitting case.

\begin{figure}
\begin{center}
\psfig{figure=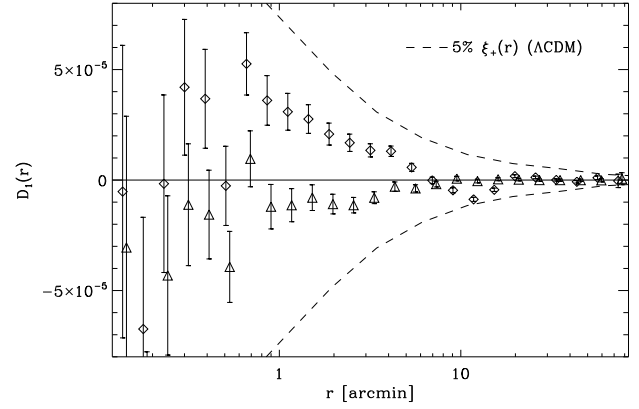,width=8.4cm,angle=0}
\caption{\label{fig:cosmolims} $D_1(r)$ fitting results for the second order (squares) and third order (triangles) chipwise fits to the starfield of Section \ref{sect:testcase} (reproduced from Figure \ref{fig:starfdiagchip}). Overlaid with the dashed line is the tolerance on $D_1(r)$ that must be met to ensure that systematic errors due to PSF variation to contribute less than $5 \%$ of the $\Lambda$CDM cosmological $\xi_+^{\gamma}$ signal for source galaxies at redshift $z=0.7$, calculated as described in Section \ref{sect:relation}.}
\end{center}
\end{figure}
The expression \eqref{eq:d1sys1} can also be reversed to define a tolerance limit upon $D_1(r)$ for a given `acceptable' level of systematics when fitting models to the spatial variation of the PSF. \citet{paulinetal08} argue that the steepness of the galaxy size distribution justifies the approximation that, typically,
\begin{equation}
\expect{\left(\frac{R_{\textrm{PSF}}}{R_{\textrm{gal}}} \right)^4} \simeq 
\left[ \left(\frac{R_{\textrm{PSF}}}{R_{\textrm{gal}}} \right)_{\textrm{min}} \right]^4 \simeq (1 / 1.5)^4 ;
\end{equation}
together with $P^{\gamma} \simeq 1.84$ this allows limits to be placed upon acceptable values of $D_1(r)$, relative to the expected cosmological signal $\xi^{\gamma}_+(r)$. Using these values, Figure \ref{fig:cosmolims} shows the tolerance upon $D_1(r)$ that must be met to ensure that anisotropy systematics do not exceed $5\%$ of the $\Lambda$CDM cosmological signal on any scale for a population of lensing source galaxies at $z=0.7$ (a typical median value for current and planned wide-area lensing surveys such as CFHTLS-W and Pan-STARRS: e.g.\ \citealp{fuetal08,hoekstraetal06,kaiser04}).  This prediction was calculated using the \citet{smithetal03} non-linear matter power spectrum for a flat, $\Lambda$CDM cosmology with $\Omega_{\textrm{m}} = 0.25$, $\Omega_{\Lambda} = 0.75$, $h=0.7$, power-law spectral index $n_{\textrm{s}}=1$ and normalization $\sigma_8 = 0.8$. Over-plotted are the second and third order chipwise fitting results of Section \ref{sect:testcase} and Figure \ref{fig:starfdiagchip}.

The arguments presented here are relatively simplistic, and the impact of residual PSF anisotropy variation upon shear estimation will also generally depend upon the correction method used. However, the approximate tolerance limits they place upon the quality of PSF modelling are a useful first guide, and the accuracy of these limits could easily be improved further by simulating erroneous PSF correction through any given shear measurement pipeline.

\section{General modelling applications}\label{sect:extension}
\begin{figure*}
\begin{center}
\psfig{figure=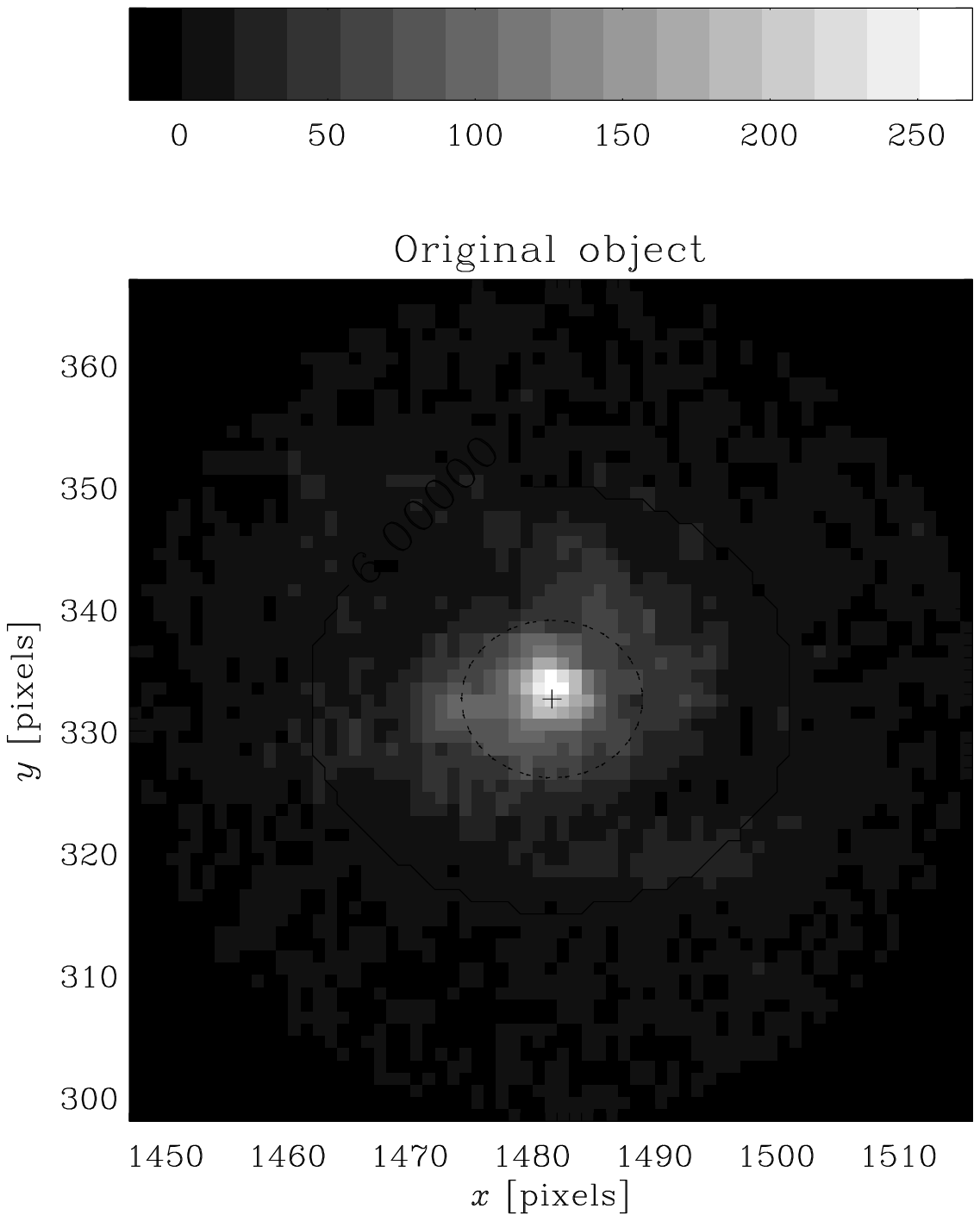,width=5.7cm,angle=0}
\psfig{figure=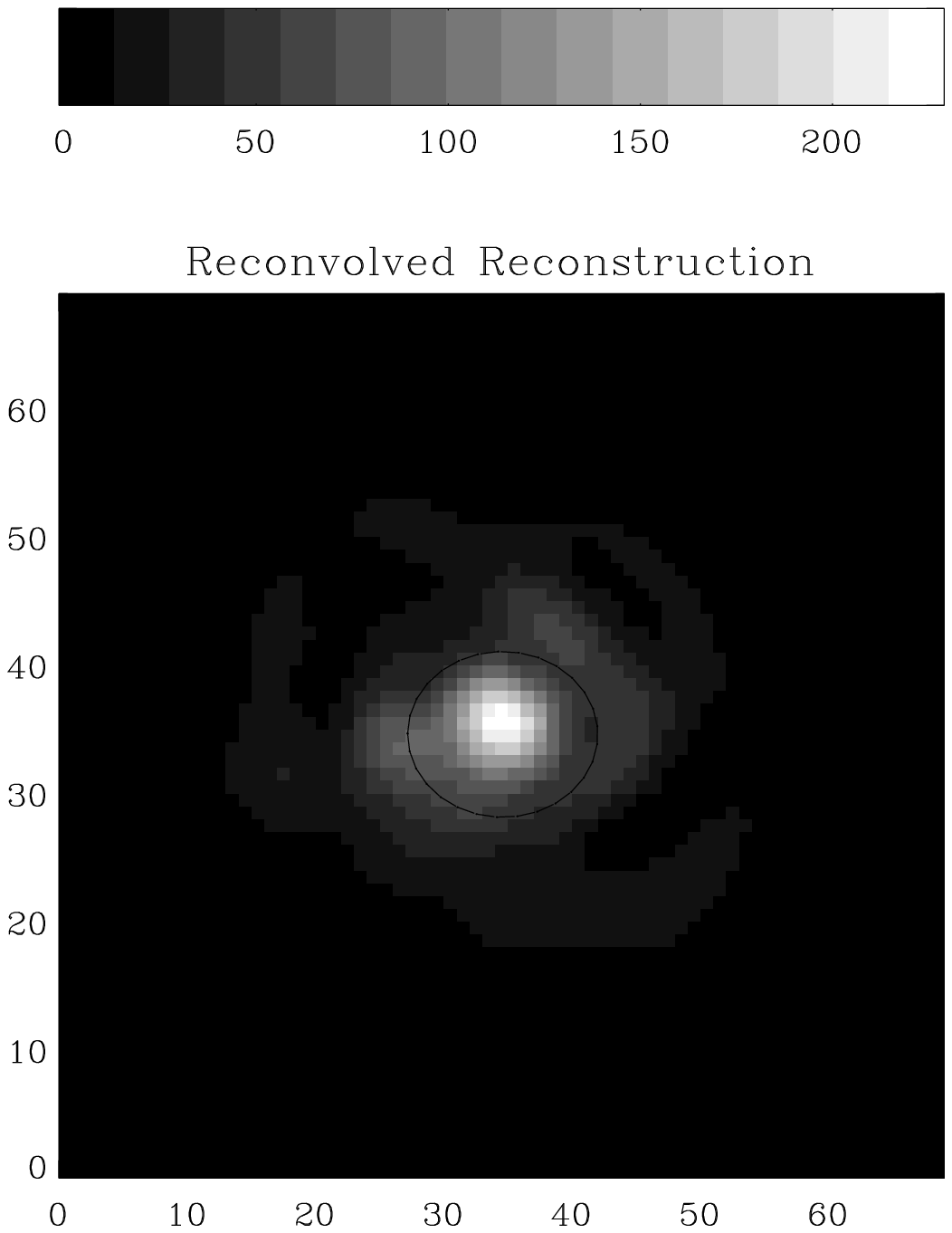,width=5.7cm,angle=0}
\psfig{figure=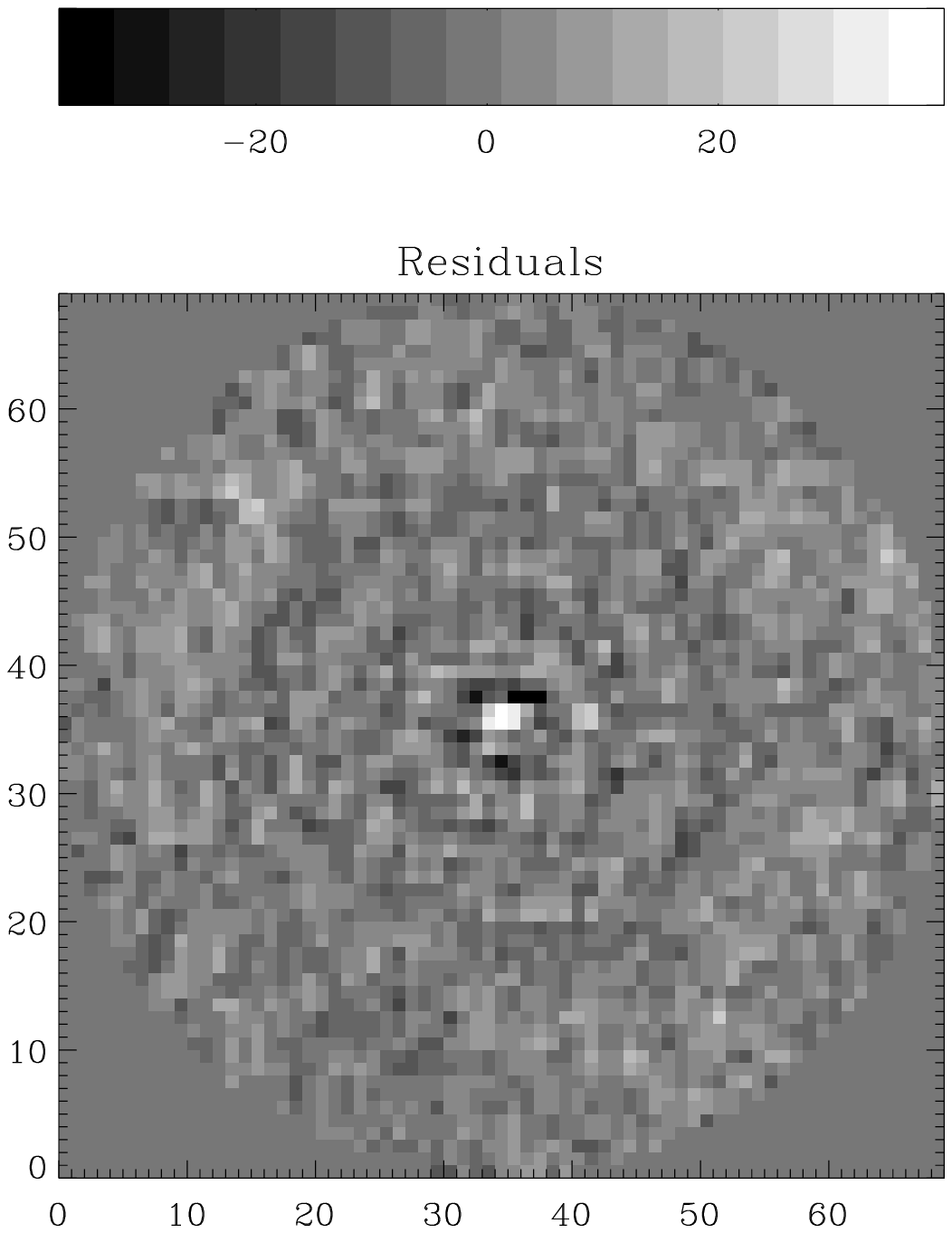,width=5.7cm,angle=0}
\caption{\label{fig:shapelets} Galaxy image, model and residuals map for a galaxy from the F606W filter \emph{Hubble} Ultra Deep Field \citep{beckwithetal06}. The image model and plots were made using the publicly-available \emph{Shapelets} software \citep{masseyrefregier05}.
In this example the galaxy shape model passes through the
    automatic shapelet modelling process with no error flags, 
and a reasonable reduced
    $\chi^2$ of 1.02087 (this is calculated using the individual pixel
    values, treating each pixel as independent and giving a total number of degrees of freedom of 3657). }
\end{center}
\end{figure*}
A final question is whether the technique can be extended
into spheres beyond the characterization of anisotropic PSFs for weak
lensing.  Another possible application of the analysis, also within
the weak lensing context, is illustrated by Figure
\ref{fig:shapelets}.  Here a shapelet model of a galaxy, imaged in the
F606W band of the \emph{Hubble Space Telescope} Ultra Deep Field
\citep{beckwithetal06}, can be seen compared to the original image and
to the map of residuals between the two; the shapelet decomposition
was performed using the publicly available code described by
\citet{masseyrefregier05}. The residual map clearly shows correlated
features but, despite this, the galaxy image passed through the
automated, iterative fitting routine of \citet{masseyrefregier05}
without any error error flags. Moreover, the model passed with a
reasonable reduced chi-squared of $\chi^2 = 1.02087$ as compared to a
good fit-expected value of $\chi^2 \simeq 1 \pm 0.0165$, calculated
via the 3657 degrees of freedom in the model (number of pixels minus
number of model coefficients).  In fact, this low value of chi-squared
may be sensitive to an erroneous overestimation in the automatic
calculation of the background sky noise used by the shapelet software,
but this simply highlights an important limitation of such techniques
when the uncertainty in the data is but imperfectly known.  The
diagnostic tools presented here are less sensitive to ignorance about
the uncertainty on individual measurements, and may be used even in
the total absence of such knowledge.

Thus motivated, a simple generalization of the technique into other
potential applications is now discussed in brief.  The correlation of
two complex functions $f$ and $g$ in an $n$-dimensional space can be
defined as
\begin{equation}\label{eq:fstarg}
[f \star g](\rb) \equiv \int_{\mathbb{R}} f^*(\xb) g(\xb + \rb) \dif^n x.
\end{equation}
In the case of the correlation functions defined in Section
\ref{sec:basic}, $f$ and $g$ would be formed from simple combinations
of the $e_{\mtan}$ and $e_{\times}$ quantities defined by equation
\eqref{eq:tanrot}.  Once more, a general function $g(\xb)$ may be
defined as a field that describes a number of discrete, noisy
samplings of an underlying `true' field $g_{\mt}(\xb)$; this
quasi-field represents the measurements:
\begin{equation}
g(\xb) = g_{\mt}(\xb) + N(\xb).
\end{equation}
As before, $N$ will be assumed to be stochastic noise.
Similarly, a model fit to these measurements can be written as
\begin{equation}
g_{\mm}(\xb) = g_{\mt}(\xb) + m(\xb, g_{\mt}, N ; f_{\mm})
\end{equation}
where $m$ is again the inaccuracy and $f_{\mm}$ labels the modelling scheme chosen to represent $g(\xb)$.

As in Section \ref{sec:basic}, a starting point is to then make simple assumptions about the properties of the noise for the data being considered, namely that
\begin{equation}\label{eq:noise1}
[N \star N](\rb) = \int_{\mathbb{R}} N^*(\xb) N(\xb + \rb) \dif^n x = 0
\end{equation}
and
\begin{equation}\label{eq:noise2}
\left[g_{\mt} \star N + N \star g_{\mt} \right](\rb) = 0.
\end{equation}
Given these assumptions, and in direct analogy to the functions defined in Section \ref{sec:diagnostics}, the following two diagnostic functions can be defined in general:
\begin{eqnarray}
D_1(\rb)  &\equiv & \left[ (g-g_{\mm}) \star (g-g_{\mm}) \right] (\rb) \\ 
     & = & \left[m \star m \right](\rb) - \left[m \star N + N \star m \right](\rb), \\
D_2(\rb)  &\equiv & \left[g \star (g-g_{\mm}) + (g-g_{\mm}) \star g \right] (\rb) \\ 
     & = & - \left[m \star g_{\mt} + g_{\mt} \star m \right](\rb) \nn 
    & ~ & - \left[m \star N + N \star m \right](\rb).
\end{eqnarray}
These functions should tend to zero for all vectors $\rb$ if the model
fit employed is both stable and accurate, and will have predictable
behaviour for over- and underfitting models in a way directly
analogous to the results of Sections \ref{sec:under} \&
\ref{sec:over}. If the noise assumptions of equations
\eqref{eq:noise1} and \eqref{eq:noise2} are broken then the functions
above will be expected to be non-zero, but tending towards a form that
may be derived from observations of pure noise.

The use of $D_1(r)$ and $D_2(r)$ may provide valuable clues to ways in
which the modelling of galaxy images such as Figure
\ref{fig:shapelets} may be improved; one immediate example might be in
the quantified selection of a more appropriate basis set than the
Gauss-Hermite polynomials used by the shapelet method
\citep{nganetal08}.  However, there is no reason why the method must
be restricted to weak lensing analyses, as the quantification and
suppression of correlated residuals is surely a justified concern
wherever physical data is being described by a best fitting model.
Correlations in residuals between neighbouring data points should be
examined wherever best-fitting models are being used to represent
physical data, including but not limited to CMB temperature power
spectra, supernova distance-redshift curves, or determinations of the
matter power spectrum from galaxy clustering.  However, such matters
are clearly a topic for further investigation; the overall findings
and results of this work will now be discussed.

\section{Conclusions}\label{sect:conc}

A simple theoretical framework for an interpretation of statistically
correlated modelling residuals has been presented, and developed into
a pair of independent two-point correlation function diagnostics for
the specific application of PSF modelling in weak lensing.  The
$D_1(r)$ and $D_2(r)$ diagnostic functions have been defined as the
autocorrelation in residuals and the data-residual cross-correlation,
which should tend to zero on all scales for good models. Visual
inspection of these functions has been shown to give sensitive insight
into model selection for a single simulated starfield designed to
mimic typical telescope PSF patterns.  This visual inspection of the
diagnostics not only quantifies the success of a given best fit model,
but also provides improvements to be made in a directed fashion via
the simple distinguishing features common to both underfitting
($D_i(r)$ expected to be both positive and negative as a function of
$r$) and overfitting models ($D_i(r)$ negative only, most likely on
small scales).

The analysis was then extended to a large suite of randomly generated starfields, and simple quantifications of the extent to which $D_1(r) = 0$ were shown to be able to select appropriate orders of modelling complexity very successfully. 
The results for $D_2(r)$ were less successful at this stage, but this may be partly explained by the greater covariance expected between neighbouring bins for this correlation function. For the arbitrary $e_{\mt}$ fields of Section \ref{sect:arbitrary} the two performed more similarly, although with $D_2(r)$ still preferring slightly more complex fitting surfaces.  Nonetheless, the success of the $D_1(r)$ diagnostic in particular provides strong evidence that the method may be used as a means of improving models of the spatial variation of the PSF in weak lensing analyses. The diagnostics successfully differentiate the stability and accuracy of competing models in cases where the chosen fitting basis both can and cannot fully reproduce the underlying $e_{\mt}$.
In cases where visual inspection and full covariance calculations are possible when making modelling choices, i.e.\ if $D_1(r)$ and $D_2(r)$ need not be calculated many times, both diagnostics may give useful insight into the properties of the model and scales on which the fit can be given greater or lesser freedom. 

Such properties are clearly of great relevance to the analysis of forthcoming weak lensing surveys, and so simple arguments were given for how to relate the $D_1(r)$ diagnostic to precision requirements for the fundamental cosmic shear observable $\xi_+^{\gamma}(r)$. Finally, the start of a generalization of the technique into more wider modelling applications was discussed. This simple work gives hope that the technique might be of use in spheres outside weak lensing.

However, the description presented is basic, largely empirical, and leaves many questions
unanswered.  Placing the analysis of the $D_1$ and $D_2$ diagnostics
on a truly statistical level, by truly estimating the \emph{probability} of
a given $D_i(\rb)$ in the null hypothesis of a well-chosen model,
would be a topic for useful future investigation. The KS test analysis attempted in this respect clearly falls somewhere short of this aim. Such analysis may
inevitably be much restricted by assumptions that will need to be made
about the properties of the noise upon data and, perhaps less
tractably, the properties of the underlying physical reality
$g_{\mt}(\xb)$ that is being modelled.  Despite these difficulties, such work might
allow the technique to do more than simply rank competing models by
quantifying agreement with $D_i(\rb) = 0$, which is where the theory
currently stands.

If a more complete statistical description can be achieved, the
analysis of correlated modelling residuals may form an extremely
useful addition to existing model selection criteria based upon
calculations of the statistical likelihood, such as the AIC, BIC and
Bayesian evidence.  The technique requires little prior knowledge of
the uncertainties upon individual data points, except for the
simplifying approximation that they are random and uncorrelated,
making it potentially more robust that chi-squared when uncertainties are difficult to
estimate.  Finally, the greatest interest in the technique may lie in
where it most differs from standard model selection criteria: the fact
that it takes the relative locations of data points, as well as their data
values, directly into account. This information is valuable, and efficient ways
to make use of it may be possible.

\section{Acknowledgments}
The author would like to thank David Bacon, Henk Hoekstra, Yannick
Mellier, Tim Schrabback and Ludo van Waerbeke for useful comments and
suggestions in the period leading up to this work. Martin Kilbinger
must also be thanked for both useful discussions and for the use of
his correlation function code, and the anonymous referee
should also be thanked for their useful suggestions which significantly
added to the breadth and detail of the paper.
This work was performed in support of
the CFHTLS Systematics Collaboration (Van Waerbeke et al., in
prep.), and the author has been supported by an Experienced Researcher
Fellowship from the European Union Dark Universe through Extragalactic
Lensing (DUEL) Research Training Network.

\bsp

\bibliographystyle{mn2e}

\bibliography{btprmnras}

\begin{thebibliography}{}

\bibitem[\protect\citeauthoryear{{Arfken} \& {Weber}}{{Arfken} \&
  {Weber}}{2005}]{arfkenweber05}
{Arfken} G.~B.,  {Weber} H.~J.,  2005, {Mathematical methods for physicists 6th
  ed.}

\bibitem[\protect\citeauthoryear{{Beckwith} et~al.,}{{Beckwith}
  et~al.}{2006}]{beckwithetal06}
{Beckwith} S.~V.~W.,  et~al., 2006, \aj, 132, 1729

\bibitem[\protect\citeauthoryear{{Bernstein} \& {Jarvis}}{{Bernstein} \&
  {Jarvis}}{2002}]{bernsteinjarvis02}
{Bernstein} G.~M.,  {Jarvis} M.,  2002, \aj, 123, 583

\bibitem[\protect\citeauthoryear{{Bridle} et~al.,}{{Bridle}
  et~al.}{2008}]{bridleetal08}
{Bridle} S.,  et~al., 2008, ArXiv e-prints: astro-ph/0802.1214

\bibitem[\protect\citeauthoryear{{Crittenden}, {Natarajan}, {Pen} \&
  {Theuns}}{{Crittenden} et~al.}{2002}]{crittendenetal02}
{Crittenden} R.~G.,  {Natarajan} P.,  {Pen} U.-L.,    {Theuns} T.,  2002, \apj,
  568, 20

\bibitem[\protect\citeauthoryear{{Efstathiou}}{{Efstathiou}}{2008}]{efstathiou%
08}
{Efstathiou} G.,  2008, \mnras, 388, 1314

\bibitem[\protect\citeauthoryear{{Eifler}, {Schneider} \& {Krause}}{{Eifler}
  et~al.}{2009}]{eifleretal09}
{Eifler} T.,  {Schneider} P.,    {Krause} E.,  2009, ArXiv e-prints:
  astro-ph/0907.2320

\bibitem[\protect\citeauthoryear{{Fu} et~al.,}{{Fu}  et~al.}{2008}]{fuetal08}
{Fu} L.,  et~al., 2008, \aap, 479, 9

\bibitem[\protect\citeauthoryear{{Fu} \& {Kilbinger}}{{Fu} \&
  {Kilbinger}}{2009}]{fukilbinger09}
{Fu} L.,  {Kilbinger} M.,  2009, ArXiv e-prints: astro-ph/0907.0795

\bibitem[\protect\citeauthoryear{Gelman, Carlin, Stern \& Rubin}{Gelman
  et~al.}{2003}]{gelmanetal03}
Gelman A.,  Carlin J.~B.,  Stern H.~S.,    Rubin D.~B.,  2003, Bayesian Data
  Analysis, Second Edition.
{Chapman \& Hall/CRC}

\bibitem[\protect\citeauthoryear{{Heymans} et~al.,}{{Heymans}
  et~al.}{2005}]{heymansetal05}
{Heymans} C.,  et~al., 2005, \mnras, 361, 160

\bibitem[\protect\citeauthoryear{{Heymans} et~al.,}{{Heymans}
  et~al.}{2006}]{heymansetal06step}
{Heymans} C.,  et~al., 2006, \mnras, 368, 1323

\bibitem[\protect\citeauthoryear{{Heymans} et~al.,}{{Heymans}
  et~al.}{2008}]{heymansetal08}
{Heymans} C.,  et~al., 2008, \mnras, 385, 1431

\bibitem[\protect\citeauthoryear{{Hoekstra}}{{Hoekstra}}{2004}]{hoekstra04}
{Hoekstra} H.,  2004, \mnras, 347, 1337

\bibitem[\protect\citeauthoryear{{Hoekstra}}{{Hoekstra}}{2007}]{hoekstra07}
{Hoekstra} H.,  2007, \mnras, 379, 317

\bibitem[\protect\citeauthoryear{{Hoekstra}, {Franx}, {Kuijken} \&
  {Squires}}{{Hoekstra} et~al.}{1998}]{hoekstraetal98}
{Hoekstra} H.,  {Franx} M.,  {Kuijken} K.,    {Squires} G.,  1998, \apj, 504,
  636

\bibitem[\protect\citeauthoryear{{Hoekstra}, {Hsieh}, {Yee}, {Lin} \&
  {Gladders}}{{Hoekstra} et~al.}{2005}]{hoekstraetal05}
{Hoekstra} H.,  {Hsieh} B.~C.,  {Yee} H.~K.~C.,  {Lin} H.,    {Gladders} M.~D.,
   2005, \apj, 635, 73

\bibitem[\protect\citeauthoryear{{Hoekstra}, {Mellier}, {Van Waerbeke},
  {Semboloni}, {Fu}, {Hudson}, {Parker}, {Tereno} \& {Benabed}}{{Hoekstra}
  et~al.}{2006}]{hoekstraetal06}
{Hoekstra} H.,  {Mellier} Y.,  {Van Waerbeke} L.,  {Semboloni} E.,  {Fu} L.,
  {Hudson} M.~J.,  {Parker} L.~C.,  {Tereno} I.,    {Benabed} K.,  2006, \apj,
  647, 116

\bibitem[\protect\citeauthoryear{{Jarvis} \& {Jain}}{{Jarvis} \&
  {Jain}}{2004}]{jarvisjain04}
{Jarvis} M.,  {Jain} B.,  2004, ArXiv e-prints: astro-ph/0412234

\bibitem[\protect\citeauthoryear{{Jarvis}, {Schechter} \& {Jain}}{{Jarvis}
  et~al.}{2008}]{jarvisetal08}
{Jarvis} M.,  {Schechter} P.,    {Jain} B.,  2008, ArXiv e-prints:
  astro-ph/0810.0027

\bibitem[\protect\citeauthoryear{{Kaiser}}{{Kaiser}}{2004}]{kaiser04}
{Kaiser} N.,  2004, in {J.~M.~Oschmann Jr.} ed., Society of Photo-Optical
  Instrumentation Engineers (SPIE) Conference Series Vol.~5489 of Presented at
  the Society of Photo-Optical Instrumentation Engineers (SPIE) Conference,
  {Pan-STARRS: a wide-field optical survey telescope array}.
pp 11--22

\bibitem[\protect\citeauthoryear{{Kaiser}, {Squires} \& {Broadhurst}}{{Kaiser}
  et~al.}{1995}]{kaiseretal95}
{Kaiser} N.,  {Squires} G.,    {Broadhurst} T.,  1995, \apj, 449, 460

\bibitem[\protect\citeauthoryear{{Kurek} \& {Szyd{\l}owski}}{{Kurek} \&
  {Szyd{\l}owski}}{2008}]{kurekszydlowski08}
{Kurek} A.,  {Szyd{\l}owski} M.,  2008, \apj, 675, 1

\bibitem[\protect\citeauthoryear{{Leauthaud} et~al.,}{{Leauthaud}
  et~al.}{2007}]{leauthaudetal07}
{Leauthaud} A.,  et~al., 2007, \apjs, 172, 219

\bibitem[\protect\citeauthoryear{{Liddle}, {Mukherjee} \& {Parkinson}}{{Liddle}
  et~al.}{2006}]{liddleetal06}
{Liddle} A.,  {Mukherjee} P.,    {Parkinson} D.,  2006, Astronomy and
  Geophysics, 47, 040000

\bibitem[\protect\citeauthoryear{{Liddle}}{{Liddle}}{2007}]{liddle07}
{Liddle} A.~R.,  2007, \mnras, 377, L74

\bibitem[\protect\citeauthoryear{{Lupton}}{{Lupton}}{1993}]{lupton93}
{Lupton} R.,  1993, {Statistics in theory and practice}.
Princeton, N.J.: Princeton University Press, |c1993

\bibitem[\protect\citeauthoryear{{Mandelbaum}, {Seljak}, {Kauffmann}, {Hirata}
  \& {Brinkmann}}{{Mandelbaum} et~al.}{2006}]{mandelbaumetal06gglens}
{Mandelbaum} R.,  {Seljak} U.,  {Kauffmann} G.,  {Hirata} C.~M.,    {Brinkmann}
  J.,  2006, \mnras, 368, 715

\bibitem[\protect\citeauthoryear{{Massey} et~al.,}{{Massey}
  et~al.}{2007a}]{masseyetal07nature}
{Massey} R.,  et~al., 2007a, \nat, 445, 286

\bibitem[\protect\citeauthoryear{{Massey} et~al.,}{{Massey}
  et~al.}{2007b}]{masseyetal07step}
{Massey} R.,  et~al., 2007b, \mnras, 376, 13

\bibitem[\protect\citeauthoryear{{Massey} \& {Refregier}}{{Massey} \&
  {Refregier}}{2005}]{masseyrefregier05}
{Massey} R.,  {Refregier} A.,  2005, \mnras, 363, 197

\bibitem[\protect\citeauthoryear{{Ngan}, {Van Waerbeke}, {Mahdavi}, {Heymans}
  \& {Hoekstra}}{{Ngan} et~al.}{2008}]{nganetal08}
{Ngan} W.-H.~W.,  {Van Waerbeke} L.,  {Mahdavi} A.,  {Heymans} C.,
  {Hoekstra} H.,  2008, ArXiv e-prints: astro-ph/0809.3465

\bibitem[\protect\citeauthoryear{{Paulin-Henriksson}, {Amara}, {Voigt},
  {Refregier} \& {Bridle}}{{Paulin-Henriksson} et~al.}{2008}]{paulinetal08}
{Paulin-Henriksson} S.,  {Amara} A.,  {Voigt} L.,  {Refregier} A.,    {Bridle}
  S.~L.,  2008, \aap, 484, 67

\bibitem[\protect\citeauthoryear{{Paulin-Henriksson}, {Refregier} \&
  {Amara}}{{Paulin-Henriksson} et~al.}{2009}]{paulinetal09}
{Paulin-Henriksson} S.,  {Refregier} A.,    {Amara} A.,  2009, \aap, 500, 647

\bibitem[\protect\citeauthoryear{{Press}, {Teukolsky}, {Vetterling} \&
  {Flannery}}{{Press} et~al.}{1992}]{pressetal92}
{Press} W.~H.,  {Teukolsky} S.~A.,  {Vetterling} W.~T.,    {Flannery} B.~P.,
  1992, {Numerical recipes in FORTRAN. The art of scientific computing}.
Cambridge: University Press, |c1992, 2nd ed.

\bibitem[\protect\citeauthoryear{{Refregier}}{{Refregier}}{2003}]{refregier03}
{Refregier} A.,  2003, \mnras, 338, 35

\bibitem[\protect\citeauthoryear{{Refregier} \& {Bacon}}{{Refregier} \&
  {Bacon}}{2003}]{refregierbacon03}
{Refregier} A.,  {Bacon} D.,  2003, \mnras, 338, 48

\bibitem[\protect\citeauthoryear{{Rhodes}, {Massey}, {Albert}, {Collins},
  {Ellis}, {Heymans}, {Gardner}, {Kneib}, {Koekemoer}, {Leauthaud}, {Mellier},
  {Refregier}, {Taylor} \& {Van Waerbeke}}{{Rhodes}
  et~al.}{2007}]{rhodesetal07}
{Rhodes} J.~D.,  {Massey} R.~J.,  {Albert} J.,  {Collins} N.,  {Ellis} R.~S.,
  {Heymans} C.,  {Gardner} J.~P.,  {Kneib} J.-P.,  {Koekemoer} A.,  {Leauthaud}
  A.,  {Mellier} Y.,  {Refregier} A.,  {Taylor} J.~E.,    {Van Waerbeke} L.,
  2007, \apjs, 172, 203

\bibitem[\protect\citeauthoryear{{Sasiela}}{{Sasiela}}{1994}]{sasiela94}
{Sasiela} R.~J.,  1994, {Electromagnetic wave propagation in turbulence.
  Evaluation and application of Mellin transforms}

\bibitem[\protect\citeauthoryear{{Schneider}}{{Schneider}}{2006}]{schneider06}
{Schneider} P.,  2006, in {Meylan} G.,  {Jetzer} P.,  {North} P.,  {Schneider}
  P.,  {Kochanek} C.~S.,   {Wambsganss} J.,  eds, Saas-Fee Advanced Course 33:
  Gravitational Lensing: Strong, Weak and Micro {Part 3: Weak gravitational
  lensing}.
pp 269--451

\bibitem[\protect\citeauthoryear{{Schneider} \& {Kilbinger}}{{Schneider} \&
  {Kilbinger}}{2007}]{schneiderkilbinger07}
{Schneider} P.,  {Kilbinger} M.,  2007, \aap, 462, 841

\bibitem[\protect\citeauthoryear{{Schneider}, {Van Waerbeke}, {Kilbinger} \&
  {Mellier}}{{Schneider} et~al.}{2002}]{schneideretal02p1}
{Schneider} P.,  {Van Waerbeke} L.,  {Kilbinger} M.,    {Mellier} Y.,  2002,
  \aap, 396, 1

\bibitem[\protect\citeauthoryear{{Schneider}, {Van Waerbeke} \&
  {Mellier}}{{Schneider} et~al.}{2002}]{schneideretal02bmode}
{Schneider} P.,  {Van Waerbeke} L.,    {Mellier} Y.,  2002, \aap, 389, 729

\bibitem[\protect\citeauthoryear{{Schrabback}, {Erben}, {Simon}, {Miralles},
  {Schneider}, {Heymans}, {Eifler}, {Fosbury}, {Freudling}, {Hetterscheidt},
  {Hildebrandt} \& {Pirzkal}}{{Schrabback} et~al.}{2007}]{schrabbacketal07}
{Schrabback} T.,  {Erben} T.,  {Simon} P.,  {Miralles} J.-M.,  {Schneider} P.,
  {Heymans} C.,  {Eifler} T.,  {Fosbury} R.~A.~E.,  {Freudling} W.,
  {Hetterscheidt} M.,  {Hildebrandt} H.,    {Pirzkal} N.,  2007, \aap, 468, 823

\bibitem[\protect\citeauthoryear{{Schrabback} et~al.,}{{Schrabback}
  et~al.}{2009}]{schrabbacketal09}
{Schrabback} T.,  et~al., 2009, ArXiv e-prints: astro-ph/0911.0053

\bibitem[\protect\citeauthoryear{{Smith}, {Peacock}, {Jenkins}, {White},
  {Frenk}, {Pearce}, {Thomas}, {Efstathiou} \& {Couchman}}{{Smith}
  et~al.}{2003}]{smithetal03}
{Smith} R.~E.,  {Peacock} J.~A.,  {Jenkins} A.,  {White} S.~D.~M.,  {Frenk}
  C.~S.,  {Pearce} F.~R.,  {Thomas} P.~A.,  {Efstathiou} G.,    {Couchman}
  H.~M.~P.,  2003, \mnras, 341, 1311

\bibitem[\protect\citeauthoryear{{Spergel} et~al.,}{{Spergel}
  et~al.}{2007}]{spergeletal07}
{Spergel} D.~N.,  et~al., 2007, \apjs, 170, 377

\bibitem[\protect\citeauthoryear{{Tian}, {Hoekstra} \& {Zhao}}{{Tian}
  et~al.}{2009}]{tianetal09}
{Tian} L.,  {Hoekstra} H.,    {Zhao} H.,  2009, \mnras, 393, 885

\bibitem[\protect\citeauthoryear{{Trotta}}{{Trotta}}{2008}]{trotta08}
{Trotta} R.,  2008, Contemporary Physics, 49, 71

\bibitem[\protect\citeauthoryear{{Van Waerbeke}, {Mellier} \& {Hoekstra}}{{Van
  Waerbeke} et~al.}{2005}]{vanwaerbekeetal05}
{Van Waerbeke} L.,  {Mellier} Y.,    {Hoekstra} H.,  2005, \aap, 429, 75

\end{thebibliography}

\end{document}